\documentclass[11pt]{article}

\ifdefined\FASTCOMPILE
  \PassOptionsToPackage{draft}{graphicx}
\fi

\usepackage[a4paper,margin=1in]{geometry}
\usepackage{amsmath,amssymb,amsthm}
\usepackage{graphicx}
\usepackage{subcaption}
\usepackage[section]{placeins} 
\usepackage{booktabs}
\usepackage{tabularx}
\usepackage{array}
\usepackage{siunitx}
\usepackage[dvipsnames]{xcolor}
\usepackage[colorlinks=true,linkcolor=MidnightBlue,citecolor=MidnightBlue,urlcolor=MidnightBlue]{hyperref}
\usepackage{mathtools}
\usepackage{bm}
\usepackage{microtype}
\usepackage{tikz}
\usetikzlibrary{positioning}

\graphicspath{{figures/}}

\newcommand{\Z}{\mathbb{Z}}

\newcommand{\divg}{\operatorname{div}}
\newcommand{\expect}[1]{\left\langle #1 \right\rangle}
\newcommand{\Neff}{N_{\mathrm{eff}}}
\newcommand{\HHI}{\mathrm{HHI}}

\title{\textbf{Tensor-Network Inference in a Field-Coupled XY Model for Portfolio Allocation}}
\author{%
Kartikeya Chowdhry\\
\texttt{kartikeya@gscar.in}\\
\medskip
Dr.\ G Subrahmanya V.R.K. Rao\\
\texttt{subrahmanya@gscar.in}\\
\medskip
Gopal K. Saxena Centre for Advanced Research, Bangalore
}
\date{July 2026}

\hypersetup{
  pdftitle={Tensor-Network Inference in a Field-Coupled XY Model for Portfolio Allocation},
  pdfauthor={Kartikeya Chowdhry and Dr. G Subrahmanya V.R.K. Rao},
  pdfsubject={Continuous-spin network scoring for long-only portfolios},
  pdfkeywords={portfolio allocation, XY model, tensor network, correlation network, message passing}
}

\begin{document}
\maketitle

\begin{abstract}
We use tensor-network methods to evaluate a field-coupled XY model for long-only portfolio
allocation. Empirical return signals define local fields, and correlations define pairwise
couplings on a sparse path obtained from Ward clustering. Four-point Gromov hyperbolicity
characterises the tree-likeness of the underlying correlation geometry. A Fourier--Bessel
expansion expresses the continuous-spin partition function as an integer-current tensor network,
which becomes finite after current truncation. Bottom-up contraction and top-down environment
sweeps yield one-site marginals and equilibrium cosine scores; a subsequent softmax map produces
positive, fully invested weights. Across five equity markets, we examine the dependence of these
allocations on the inverse temperature \(\beta\) and concentration parameter \(\gamma\), and
compare representative portfolios with long-only Markowitz frontiers and standard benchmarks.
The same-sample results establish a tractable connection between financial time series,
continuous-spin inference, and portfolio allocation. Predictive trading performance remains
outside the scope of the study.
\end{abstract}

\par

\section{Introduction}

Tensor networks express partition functions and observables as contractions of local tensors,
providing a common language for statistical mechanics and graphical-model inference
\cite{baxter1982,nishino1996,nishino1997,levin2007,xie2012,orus2014,evenbly2015,ran2020}.
For classical XY models, Fourier character expansions replace angular variables with integer
modes, as demonstrated in studies of frustrated spin systems
\cite{song2022ffxy,song2023kagome,song2023unified}. The cost of contraction depends on the
interaction graph: loops can make exact evaluation expensive, whereas a tree admits sum-product
message passing \cite{pearl1988,kschischang2001,yedidia2003,loeliger2004,yedidia2005}.
Here we apply this framework to finance, using a field-coupled XY model on a loop-free path to
infer asset scores for long-only portfolio construction.

Figure~\ref{fig:workflow} summarises the calculation from prices to portfolio comparisons. The
correlation-distance geometry defines the Ward hierarchy and interaction path. Gibbs inference
then takes place in the separate space of XY angles.

\begin{figure}[htbp]
\centering
\begin{tikzpicture}[
  box/.style={
    draw,
    rounded corners=2pt,
    align=center,
    text width=2.85cm,
    minimum height=1.0cm,
    inner sep=3pt,
    font=\scriptsize
  },
  data/.style={box,fill=MidnightBlue!10},
  geom/.style={box,fill=ForestGreen!11},
  model/.style={box,fill=BrickRed!10},
  output/.style={box,fill=Goldenrod!13},
  arrow/.style={->,>=stealth,line width=0.65pt},
  x=1cm,y=1cm
]
  \node[data]   (n1)  at (0,0)       {Adjusted-close\\price histories};
  \node[data]   (n2)  at (3.8,0)     {Log returns, means,\\volatilities, covariance};
  \node[geom]   (n3)  at (7.6,0)     {Correlations, local fields\\and pairwise couplings};
  \node[geom]   (n4)  at (11.4,0)    {Correlation distance and\\four-point Gromov test};

  \node[geom]   (n5)  at (11.4,-1.8) {Ward hierarchy and\\sparse interaction path};
  \node[model]  (n6)  at (7.6,-1.8)  {Field-coupled XY\\Hamiltonian and Gibbs law};
  \node[model]  (n7)  at (3.8,-1.8)  {Fourier--Bessel expansion\\and current truncation};
  \node[model]  (n8)  at (0,-1.8)    {Finite tree tensor\\network};

  \node[model]  (n9)  at (0,-3.6)    {Bottom-up partition\\contraction};
  \node[model]  (n10) at (3.8,-3.6)  {Top-down environments\\and one-site marginals};
  \node[output] (n11) at (7.6,-3.6)  {Cosine scores and\\softmax portfolio weights};
  \node[output] (n12) at (11.4,-3.6) {Parameter sweeps, weight\\profiles and benchmarks};

  \draw[arrow] (n1)--(n2);
  \draw[arrow] (n2)--(n3);
  \draw[arrow] (n3)--(n4);
  \draw[arrow] (n4)--(n5);
  \draw[arrow] (n5)--(n6);
  \draw[arrow] (n6)--(n7);
  \draw[arrow] (n7)--(n8);
  \draw[arrow] (n8)--(n9);
  \draw[arrow] (n9)--(n10);
  \draw[arrow] (n10)--(n11);
  \draw[arrow] (n11)--(n12);
\end{tikzpicture}
\caption{Workflow from market data to network-adjusted long-only portfolios. The arrows show the
computational sequence; graph sparsification and current truncation are separate approximations.}
\label{fig:workflow}
\end{figure}

Correlations enter twice: they define the distance \(d_{ij}=\sqrt{2(1-\hat\rho_{ij})}\) and
the candidate interaction strengths. Four-point hyperbolicity diagnoses the resulting metric;
Ward clustering supplies a deterministic leaf order; and consecutive leaves define the loop-free
path. The geometric diagnostic and the choice of interaction graph remain separate modelling
steps.

The financial problem is to convert estimates of return and dependence into fully invested
weights \(\mathbf{w}\), with \(w_i\geq0\) and \(\sum_iw_i=1\). The classical reference is
mean--variance optimisation \cite{markowitz1952,markowitz1959}, for which
\begin{equation}
 \mathbb{E}[R_p]=\boldsymbol{\mu}^{\top}\mathbf{w},
 \qquad
 \operatorname{Var}(R_p)=\mathbf{w}^{\top}\boldsymbol{\Sigma}\mathbf{w}.
\end{equation}
Here \(\boldsymbol{\mu}=(\mu_1,\ldots,\mu_N)^\top\) is the expected-return vector and
\(\boldsymbol{\Sigma}=[\Sigma_{ij}]_{i,j=1}^N\) is the covariance matrix. These quantities
must be estimated from finite, noisy samples. Consequently, small estimation errors can produce
unstable or concentrated allocations even when the optimisation itself is accurate. Shrinkage,
regularisation, risk-based allocation, and hierarchical methods address different aspects of this
problem \cite{ledoit2004,demiguel2009,maillard2010,lopezdeprado2016}.
The method shares its empirical inputs with several established approaches to allocation.
Mean--variance theory balances expected return against variance to define an efficient frontier
\cite{markowitz1952,markowitz1959,markowitz1991}. Sensitivity to estimated moments motivates
shrinkage, constraints, minimum-variance portfolios, and equal weighting
\cite{jagannathan2003,ledoit2004,clarke2006,demiguel2009}. Risk parity controls contributions to
total risk \cite{maillard2010,clarke2013,spinu2013}, while hierarchical methods cluster assets
through correlation distance and allocate along a tree or quasi-tree
\cite{lopezdeprado2016,raffinot2017,raffinot2018,pfitzinger2019}. Here the Ward hierarchy orders
the assets, but the scores are obtained from XY inference rather than a quadratic programme or
recursive hierarchical allocation.

Statistical-mechanical formulations offer another way to organise financial decisions. Ising and
quadratic unconstrained binary optimisation (QUBO) models are natural for asset inclusion and
cardinality constraints \cite{lucas2014,rosenberg2016,glover2019,venturelli2019,mugel2022}.
Representing a continuous allocation, however, requires a finite number of precision bits, for
example
\begin{equation}
 w_i\approx\sum_{k=1}^{p}2^{-k}x_{ik},
 \qquad x_{ik}\in\{0,1\}.
\end{equation}
The number of logical variables therefore grows with both the asset universe and the requested
weight resolution. These precision bits describe the allocation itself; ancillary variables for
constraint reduction or hardware embedding serve a separate purpose. We instead place
asset-specific signals and empirical dependence in a continuous-state graphical model, without
prescribing a binary grid for capital weights.

The connection with binary optimisation is also methodological. Ising/QUBO formulations
\cite{lucas2014,glover2019} have been used for portfolio selection, trading trajectories, and
dynamic allocation with quantum annealing and quantum-inspired methods
\cite{rosenberg2016,venturelli2019,mugel2022}. These formulations directly represent discrete
decisions, while continuous weights require finite binary precision. Our model retains local
fields and interacting states but replaces binary spins with continuous XY angles; capital
weights satisfying the simplex constraints are assigned after Gibbs inference.

Each asset carries an auxiliary XY angle \(\theta_i\in[0,2\pi)\), with Hamiltonian
\begin{equation}
 H(\boldsymbol{\theta})=
 -\sum_{(i,j)\in E}J_{ij}\cos(\theta_i-\theta_j)-\sum_i h_i\cos\theta_i .
\end{equation}
The edge set \(E\) specifies the chosen interaction graph, with \((i,j)\in E\) denoting an
asset pair coupled by \(J_{ij}\). Correlations determine the couplings \(J_{ij}\), and local
return signals determine the fields \(h_i\). Two opposite angular orientations recover the
Ising state space; allowing the full circle gives one continuous latent variable per asset.
The angle \(\theta_i\) is an auxiliary state whose thermally averaged projection supplies a
network score. Capital weights are assigned after inference. Positive correlations promote
alignment between neighbouring states, while covariance risk enters explicitly in the Markowitz
comparison. The spin construction thus defines a probabilistic scoring model without identifying
financial markets with physical magnetic systems.

In this sense, the construction connects statistical mechanics, network smoothing, and portfolio
allocation. Return and dependence estimates determine the fields and couplings, the character
expansion enables marginal inference, and the marginals determine portfolio scores. Positive
correlation couplings transmit coherent signals between neighbouring latent states. They do not
penalise covariance risk, so diversification is assessed empirically and against the field-only
ablation. The hierarchy supplies an ordering rather than an allocation rule.

The construction involves two distinct geometries. Assets are compared through the correlation
distance \(d_{ij}=\sqrt{2(1-\widehat{\rho}_{ij})}\), whereas the spins take values on \(S^1\).
For each asset quadruple, the four-point Gromov test compares the two largest sums of opposite-pair
distances. Half their difference is zero for an exact tree metric and measures departures from
tree-likeness \cite{gromov1987,bridson1999,jonckheere2011,chen2012}. This diagnostic acts directly
on the empirical distance matrix, requires no prespecified tree, and retains the exact tree-metric
condition as its reference.

Ward clustering then supplies a hierarchy whose leaf order defines the sparse interaction path
\cite{ward1963}. Its minimum-variance criterion is applicable because correlation distance is
Euclidean for standardised returns. By favouring compact clusters and reducing the chaining
associated with single linkage, Ward ordering places related assets near one another before
consecutive leaves are connected. Complete or average linkage would generally produce different
paths; we keep Ward linkage fixed and do not compare alternatives. The clustering hierarchy and
the interaction graph have different roles: only consecutive leaf pairs interact. Hyperbolicity
neither selects this path nor bounds the error of sparsification. Restricting the integer currents
in the subsequent tensor-network calculation introduces a separate numerical approximation.

For the resulting field-coupled model, the partition function is
\begin{equation*}
 Z=\prod_{k\in V}\int_0^{2\pi}\frac{d\theta_k}{2\pi}
 \exp[-\beta H(\boldsymbol{\theta})] .
\end{equation*}
Applying the Fourier--Bessel identity
\begin{equation}
 e^{x\cos\theta}=\sum_{n\in\mathbb{Z}}I_n(x)e^{in\theta}
\end{equation}
to every interaction and field factor replaces the angular integrals with a tensor network of
integer currents \(n\), weighted by modified Bessel functions \(I_n\) of the first kind.
We follow the character-expansion approach used for classical XY models by F.-F. Song and
co-workers \cite{song2022ffxy,song2023kagome,song2023unified}. Restricting the infinite current
sum to \(|n|\leq K\) gives finite bond dimensions and permits numerical contraction. The
resulting cutoff error is additional to the modelling error introduced when the dense correlation
graph is replaced by the Ward-ordered path.

On a rooted graph, a bottom-up contraction from the leaves gives the truncated partition
function. A top-down sweep then supplies the environments needed for one-site marginals and
equilibrium scores \(m_i=\langle\cos\theta_i\rangle\). A softmax converts these scores to
capital weights,
\begin{equation}
 w_i=\frac{\exp(\gamma m_i)}{\sum_j\exp(\gamma m_j)} .
\end{equation}
The inverse temperature \(\beta\) sets the response of the Gibbs distribution to fields and
couplings. The parameter \(\gamma\) acts after inference and controls allocation concentration:
\(\gamma=0\) gives equal weights, and larger \(\gamma\) favours assets with larger scores.
Across \((\beta,\gamma)\), we study the marginals, the scores \(m_i\), and the resulting
weights in five equity markets. The contractions produce normalised marginals and positive,
fully invested allocations, with market-dependent score and concentration patterns. Comparisons
with Markowitz frontiers and standard benchmarks use the same estimation sample, establishing
the behaviour and computational feasibility of the construction. Predictive performance,
statistical superiority, and trading viability require out-of-sample evaluation.

Sections~\ref{sec:model} and~\ref{sec:geometry} specify the statistical-mechanical model and its
sparse interaction graph. Sections~\ref{sec:tensor-representation} and~\ref{sec:tensor-contraction}
derive the current representation, contraction, and observables, and Section~\ref{sec:portfolio-map}
converts those observables into portfolio weights. Section~\ref{sec:results} presents the
numerical study; Sections~\ref{sec:discussion} and~\ref{sec:conclusion} discuss its interpretation
and conclusions. General tensor-network background is retained in
Appendix~\ref{app:tensor-background}.

\section{Financial inputs and the field-coupled XY model}
\label{sec:model}

\subsection{Portfolio construction and the Markowitz formulation}
Consider a fixed amount of capital distributed among \(N\) assets. The allocation vector
\(\mathbf w=(w_1,\ldots,w_N)\) records the fraction \(w_i\) invested in each asset \(i\).
For a fully invested, long-only portfolio, the weights satisfy
\begin{equation}
    \mathbf 1^\top\mathbf w=1,
    \qquad
    w_i\geq0 .
\label{eq:simplex}
\end{equation}
Expected return alone is insufficient to assess an allocation because asset returns are uncertain
and mutually dependent. Markowitz theory \cite{markowitz1952,markowitz1959,markowitz1991}
describes these inputs through the expected-return vector \(\boldsymbol\mu\) and covariance
matrix \(\boldsymbol\Sigma\), giving
\begin{equation}
    \mathbb E[R_p]=\boldsymbol\mu^\top\mathbf w,
    \qquad
    \operatorname{Var}(R_p)=\mathbf w^\top\boldsymbol\Sigma\mathbf w .
\label{eq:markowitzmoments}
\end{equation}
A standard long-only problem is
\begin{equation}
\begin{aligned}
    \min_{\mathbf w}\quad &\mathbf w^\top\boldsymbol\Sigma\mathbf w,\\
    \text{subject to}\quad
    &\boldsymbol\mu^\top\mathbf w\geq r_\star,\qquad
      \mathbf 1^\top\mathbf w=1,\qquad \mathbf w\geq0 .
\end{aligned}
\label{eq:markowitzqp}
\end{equation}
Varying the target return \(r_\star\) traces the estimated efficient frontier. Convexity makes
the optimisation well defined for fixed inputs, but estimation noise in the means and covariances
can still make the resulting weights unstable.

\subsection{From financial time series to model coefficients}
Let \(P_{i,t}\) denote the adjusted closing price of asset \(i\) on day \(t\). Log returns are
defined as
\[
    r_{i,t}=\log P_{i,t}-\log P_{i,t-1}.
\]
The sample mean and covariance of log returns are
\[
    \hat\mu_i = \frac{1}{T}\sum_{t=1}^T r_{i,t},
    \qquad
    \hat\Sigma_{ij}
    =
    \frac{1}{T-1}\sum_{t=1}^T (r_{i,t}-\hat\mu_i)(r_{j,t}-\hat\mu_j).
\]
The corresponding sample correlation is
\[
    \hat\rho_{ij}
    =
    \frac{\hat\Sigma_{ij}}{\sqrt{\hat\Sigma_{ii}\hat\Sigma_{jj}}}.
\]
We use these statistics to define the continuous-spin model: pairwise dependence supplies the
dense couplings, and the local return signal supplies the fields,
\begin{equation}
    J_{ij}=\hat\rho_{ij},
    \qquad
    h_i=\eta s_i,
    \qquad
    s_i=\frac{\hat\mu_i}{\sqrt{\hat\Sigma_{ii}}}.
\label{eq:coefficients}
\end{equation}
All experiments use the dimensionless daily score \(s_i\) with \(\eta=1\). Positive
correlations favour alignment, and the sign of \(h_i\) favours orientations toward or away
from the reference axis. Once the graph is sparsified, \(J_{ij}\) is non-zero only on retained
edges.

\subsection{Binary allocation in QUBO and Ising form}
To relate the construction to discrete formulations, let \(x_i\in\{0,1\}\) indicate whether
asset \(i\) is selected. A continuous capital fraction may be represented by \(p\) precision
bits,
\begin{equation}
    w_i\approx\sum_{k=1}^{p}2^{-k}x_{ik},
    \qquad x_{ik}\in\{0,1\}.
\label{eq:binaryweight}
\end{equation}
Collecting all bits in \(\mathbf x\), the portfolio objective together with a budget penalty
takes the QUBO form
\begin{equation}
    \min_{\mathbf x\in\{0,1\}^{M}}
    \left[
      \mathbf x^\top Q\mathbf x+\mathbf q^\top\mathbf x
      +A(\mathbf c^\top\mathbf x-1)^2
    \right],
\label{eq:qubo}
\end{equation}
where \(A>0\). Under \(x_a=(1+\sigma_a)/2\), \(\sigma_a\in\{-1,+1\}\), the objective becomes,
up to an additive constant,
\begin{equation}
    H_{\mathrm{Ising}}(\boldsymbol\sigma)
    =
    -\sum_{a<b}\mathcal J_{ab}\sigma_a\sigma_b
    -\sum_a\mathcal h_a\sigma_a .
\label{eq:ising}
\end{equation}
This representation directly accommodates inclusion and cardinality decisions. For continuous
allocation, it requires \(Np\) logical variables and fixes the weight resolution before the
problem is solved.

\subsection{From Ising spins to XY rotors}
An Ising spin has two possible orientations, \(\sigma_i=\pm1\), along a fixed axis. An XY
spin instead takes values on the unit circle,
\begin{equation}
    \mathbf S_i=(\cos\theta_i,\sin\theta_i),
    \qquad \theta_i\in[0,2\pi).
\label{eq:xyspin}
\end{equation}
The scalar product \(\mathbf S_i\cdot\mathbf S_j=\cos(\theta_i-\theta_j)\) gives a
rotationally invariant pair interaction. Restricting \(\theta_i\) to \(0\) or \(\pi\)
recovers \(\cos(\theta_i-\theta_j)=\sigma_i\sigma_j\) exactly. The XY model is therefore
the minimal \(O(2)\) continuous-spin extension of the Ising alignment model.

Local fields orient the rotors, pairwise couplings transmit alignment or anti-alignment, and
temperature controls fluctuations about low-energy configurations. In the financial application,
\(\theta_i\) remains an auxiliary state. The thermal average of \(\cos\theta_i\) is used as
a network-adjusted score; the angle itself is not a capital fraction.

\subsection{Field-coupled XY Hamiltonian}
For a graph \(G=(V,E)\), the field-coupled XY Hamiltonian is
\begin{equation}
H(\bm\theta)
=
-\sum_{(i,j)\in E} J_{ij}\cos(\theta_i-\theta_j)
-\sum_{i\in V} h_i\cos\theta_i .
\label{eq:H}
\end{equation}
As above, restricting every angle to two opposite directions reduces Equation~\eqref{eq:H} to
an Ising Hamiltonian with pairwise spin products and local fields. The pair term couples relative
orientations through empirical dependence, while the local field \(h_i\) biases each spin according
to its own return signal. We choose the common field axis as \(\theta=0\). Although the axis is a
coordinate convention, non-zero fields break the global rotational symmetry of the zero-field
model.

\subsection{Gibbs distribution}
At inverse temperature \(\beta>0\), the model defines the Gibbs distribution
\begin{equation}
p(\bm\theta)
=
\frac{1}{Z}
\exp\left[
\beta\sum_{(i,j)\in E} J_{ij}\cos(\theta_i-\theta_j)
+\beta\sum_{i\in V} h_i\cos\theta_i
\right],
\label{eq:gibbs}
\end{equation}
where the partition function is
\begin{equation}
Z
=
\prod_{k\in V}\int_0^{2\pi}\frac{d\theta_k}{2\pi}
\exp[-\beta H(\bm\theta)] .
\label{eq:partition}
\end{equation}
All expectations in the paper are taken with respect to this distribution.

Numerical evaluation proceeds through the Fourier--Bessel character expansion. The exact
integer-current form \eqref{eq:flux}, finite-current approximation \eqref{eq:truncatedflux}, and
adaptive cutoff \eqref{eq:adaptiveK} are derived in Section~\ref{subsec:fourier-current}.
This cutoff controls the current sum on the graph chosen in Section~\ref{sec:geometry}; graph
sparsification and current truncation are separate approximations.

\subsection{Equilibrium magnetisation as an asset score}
The observable used for allocation is the equilibrium magnetisation along the reference axis,
\begin{equation}
    m_i=\expect{\cos\theta_i}.
\label{eq:mi}
\end{equation}
It depends on the local field \(h_i\), neighbouring fields, and the interaction structure of the
whole graph. Assets with similar individual return scores can therefore acquire different
magnetisations through their network positions. The response identity
\begin{equation}
  m_i=\frac{1}{\beta}\frac{\partial\log Z}{\partial h_i},
\label{eq:response}
\end{equation}
provides an independent check on the calculation. Two limits isolate the role of interactions.
When all fields vanish, reflection and rotational symmetry imply \(m_i=0\). When all couplings
vanish, the sites become independent and
\begin{equation}
  m_i^{(0)}=\frac{I_1(\beta h_i)}{I_0(\beta h_i)}.
\label{eq:fieldonly}
\end{equation}
Equation~\eqref{eq:fieldonly} provides the analytic field-only reference for the interacting model.

\subsection{Role of the inverse temperature}
The empirical calculation treats \(\beta\) as an adjustable parameter rather than estimating a
market-equilibrium temperature. Only the products \(\beta J_{ij}\) and \(\beta\eta s_i\)
enter the Gibbs law; \(\beta\), the coupling scale, and \(\eta\) are therefore not separately
identifiable without a calibration convention. We fix \(J_{ij}=\hat\rho_{ij}\) on retained
edges and \(\eta=1\), and then vary \(\beta\).

As \(\beta\to0\), the Gibbs state approaches the uniform measure on the \(N\)-torus and
\(m_i\to0\). Intermediate \(\beta\) allows the field vector and interaction network to distinguish
assets. At very large values, low-energy orientations dominate, potentially reducing sensitivity
to small differences in the inputs.

\section{Correlation geometry and the sparse interaction path}
\label{sec:geometry}

\subsection{Correlation matrices from financial time series}
Within each market, adjusted-close prices are converted to log returns
\(r_{i,t}=\log P_{i,t}-\log P_{i,t-1}\) and aligned on a complete-case calendar. Define the
standardised series by
\[
    z_{i,t}=\frac{r_{i,t}-\hat\mu_i}{\hat\sigma_i},
    \qquad
    \hat\sigma_i=\sqrt{\hat\Sigma_{ii}} .
\]
The sample correlation coefficient is the covariance of the standardised return series,
\begin{equation}
    \hat\rho_{ij}
    =
    \frac{1}{T-1}\sum_{t=1}^{T}z_{i,t}z_{j,t}
    =
    \frac{\hat\Sigma_{ij}}
    {\sqrt{\hat\Sigma_{ii}\hat\Sigma_{jj}}}.
\label{eq:corrmatrix}
\end{equation}
The pairwise coefficients form the symmetric matrix \(\widehat{\bm\rho}\), whose diagonal
entries are one.
\subsection{Correlation distance}

The empirical correlation matrix is converted to a distance matrix using
\begin{equation}
    d_{ij}=\sqrt{2(1-\hat\rho_{ij})}.
\label{eq:corrdist}
\end{equation}
This standard distance in correlation-network analysis \cite{mantegna1999} vanishes for perfectly
correlated assets and increases as correlation decreases. It is used to diagnose the geometry and
construct the sparse graph. The spin couplings on retained edges remain the empirical
correlations themselves.

\subsection{Four-point Gromov hyperbolicity}

Tree-likeness is assessed using the four-point Gromov hyperbolicity condition
\cite{gromov1987,bridson1999}. For four assets \(a,b,c,d\), define the three sums
\[
S_1=d_{ab}+d_{cd},\qquad
S_2=d_{ac}+d_{bd},\qquad
S_3=d_{ad}+d_{bc}.
\]
After ordering them as \(S_{(1)}\ge S_{(2)}\ge S_{(3)}\), set
\[
    \delta(a,b,c,d)=\frac{1}{2}\left(S_{(1)}-S_{(2)}\right).
\]
For an exact tree metric, \(\delta=0\) for every quadruple. Four points admit precisely three
pairings into disjoint pairs, and equality of the two largest pairing sums characterises tree
geometry. Fewer points cannot support this comparison. For \(n>4\), the test is applied to every
four-point subset; the largest discrepancy is the hyperbolicity constant of the finite metric.

Small \(\delta/\operatorname{diam}(d)\) indicates approximate tree-likeness in this four-point
sense \cite{jonckheere2011,chen2012,fluschnik2016,coudert2022}. The empirical statistics and their
distributions appear in Table~\ref{tab:gromov} and Figure~\ref{fig:gromov}.

\subsection{Ward clustering, interaction paths, and approximation scope}

We apply Ward hierarchical clustering \cite{ward1963} to the distance matrix
\eqref{eq:corrdist}. For complete-case samples, this distance is Euclidean between standardised
return vectors up to scale, so Ward's variance criterion applies. The resulting leaf order
\(\pi(1),\ldots,\pi(N)\) defines a nearest-neighbour interaction path with edges
\[
    (\pi(1),\pi(2)),\;(\pi(2),\pi(3)),\ldots,(\pi(N-1),\pi(N)).
\]
Each retained edge carries its empirical correlation as its coupling. The hierarchy does not
uniquely determine this path: exchanging the children of an internal dendrogram node preserves
the hierarchy but may change consecutive leaf pairs. We use the deterministic pre-order returned
by the clustering routine, without optimising over equivalent orders.

Sparsification reduces the \(N(N-1)/2\) dense-graph couplings to \(N-1\) path edges: 29 of
435 for \(N=30\), or 28 of 406 for \(N=29\). Eliminating loops reduces contraction width
while discarding most direct interactions. The three steps should therefore be distinguished.
Hyperbolicity diagnoses the correlation metric, Ward clustering selects a hierarchy, and the
consecutive-leaf rule turns its ordering into a path. Small four-point discrepancies motivate a
tree-based representation but select neither the Ward hierarchy nor a unique leaf order, and do
not bound the resulting partition-function error.

Figure~\ref{fig:dendro} displays the Ward trees for all five markets. Its vertical axis is Ward
linkage height, and the coloured cut serves only as a visual aid. The computational path follows
successive leaves in the displayed order.

\section{Tensor-network representation of the partition function}
\label{sec:tensor-representation}
\subsection{Tensor-network preliminaries}

We use the graphical notation of Biamonte and Bergholm \cite{biamonte2017}: a node denotes a
tensor, and each incident line, or \emph{leg}, denotes an index. Open legs are output indices;
joining two legs contracts their shared index. For example,
\begin{equation}
    C_{ij}=\sum_{\alpha=1}^{\chi}A_{i\alpha}B_{\alpha j}.
\label{eq:tnprimer}
\end{equation}
The size \(\chi\) of the contracted index is the \emph{bond dimension}, which controls
representational capacity and contraction cost. Contracting all legs produces a scalar such as
the partition function. Leaving a local leg open, or modifying its tensor, gives access to
marginals and observables. Appendix~\ref{app:tensor-background} reviews the standard
architectures and their relation to contraction complexity.

\subsection{Fourier--Bessel expansion and integer-current representation}
\label{subsec:fourier-current}

The Fourier--Bessel expansion converts the continuous XY integral into a discrete sum over
integer currents on the interaction edges. Expanding the edge and field factors in circular
Fourier characters introduces modified-Bessel weights. Angular integration then imposes a local
conservation constraint between incident currents and the field charge. Eliminating the field
charges leaves edge currents coupled through their vertex divergence. These currents become the
bond indices of a tensor network, whose contraction is tractable on the selected loop-free
graph. Related formulations for classical XY models are developed by F.-F. Song and co-workers
\cite{song2022ffxy,song2023kagome,song2023unified}.

Choose an arbitrary orientation for each undirected edge. Writing out Equation~\eqref{eq:partition}
gives
\begin{equation}
Z=
\prod_{k\in V}\int_0^{2\pi}\frac{d\theta_k}{2\pi}
\exp\!\left[
\beta\sum_{(i,j)\in E}J_{ij}\cos(\theta_i-\theta_j)
+\beta\sum_{i\in V}h_i\cos\theta_i
\right].
\label{eq:A1}
\end{equation}
The Fourier--Bessel character expansions are
\begin{equation}
\begin{aligned}
e^{\beta J_{ij}\cos(\theta_i-\theta_j)}
 &=\sum_{n_{ij}\in\Z}I_{n_{ij}}(\beta J_{ij})
   e^{in_{ij}(\theta_i-\theta_j)},\\
e^{\beta h_i\cos\theta_i}
 &=\sum_{q_i\in\Z}I_{q_i}(\beta h_i)e^{iq_i\theta_i}.
\end{aligned}
\label{eq:A2}
\end{equation}
The integer \(n_{ij}\) is the current on the oriented edge \(i\to j\), and \(q_i\) is
the field charge. Substituting \eqref{eq:A2} into \eqref{eq:A1} and collecting the phase at each
vertex yields
\begin{equation}
\begin{aligned}
Z={}&
\sum_{\{n_{ij}\}}\sum_{\{q_i\}}
\left[\prod_{(i,j)\in E}I_{n_{ij}}(\beta J_{ij})\right]
\left[\prod_{i\in V}I_{q_i}(\beta h_i)\right]\\
&\times
\prod_{k\in V}\int_0^{2\pi}\frac{d\theta_k}{2\pi}
e^{i[q_k+(\divg n)_k]\theta_k},
\end{aligned}
\label{eq:A3}
\end{equation}
where
\[
(\divg n)_k=
\sum_{j:(k,j)\in E}n_{kj}-\sum_{i:(i,k)\in E}n_{ik}.
\]
The orthogonality relation
\(\int_0^{2\pi}d\theta\,e^{im\theta}/(2\pi)=\delta_{m,0}\) imposes local current conservation:
\begin{equation}
Z=
\sum_{\{n_{ij}\}}\sum_{\{q_i\}}
\left[\prod_{(i,j)\in E}I_{n_{ij}}(\beta J_{ij})\right]
\left[\prod_{i\in V}I_{q_i}(\beta h_i)\right]
\prod_{k\in V}\delta_{q_k+(\divg n)_k,0}.
\label{eq:A4}
\end{equation}
Summing each constrained \(q_k\) and using \(I_{-m}(x)=I_m(x)\) yields the exact flux
representation
\begin{equation}
Z=
\sum_{\{n_{ij}\}\in\Z^E}
\prod_{(i,j)\in E}I_{n_{ij}}(\beta J_{ij})
\prod_{k\in V}I_{(\divg n)_k}(\beta h_k).
\label{eq:flux}
\end{equation}
Numerical evaluation restricts every edge current to
\(\mathcal N_K=\{-K,\ldots,K\}\):
\begin{equation}
Z_K=
\sum_{\{n_{ij}\}\in\mathcal N_K^E}
\prod_{(i,j)\in E}I_{n_{ij}}(\beta J_{ij})
\prod_{k\in V}I_{(\divg n)_k}(\beta h_k),
\qquad D=2K+1.
\label{eq:truncatedflux}
\end{equation}
The reported calculations use the conservative adaptive window
\begin{equation}
K(\beta)=\max\!\left\{12,
\left\lceil \beta J_{\max}+8\sqrt{\beta J_{\max}+1}+5\right\rceil\right\},
\qquad
J_{\max}=\max_{(i,j)\in E}|J_{ij}|.
\label{eq:adaptiveK}
\end{equation}
The exact expansion contains infinitely many positive and negative currents, whereas numerical
contraction requires a finite bond dimension. The edge argument \(\beta J_{ij}\) sets the
relevant Bessel scale, whose maximum is \(\beta J_{\max}\). At larger \(\beta\), appreciable
coefficients extend to higher current orders, so a fixed cutoff may omit significant
low-temperature contributions. Equation~\eqref{eq:adaptiveK} increases \(K\) with
\(\beta J_{\max}\), adds a square-root safety margin, and enforces a minimum window at weak
coupling. This conservative prescription is not a universal error bound. We assess its numerical
accuracy by increasing \(K\) and checking the stability of the observables.

When all fields vanish, \(I_q(0)=\delta_{q,0}\) restricts the sum to divergence-free currents.
Non-zero fields permit open currents whose endpoints carry local field weights. For negative
Bessel arguments, \(I_n(-x)=(-1)^nI_n(x)\) for integer \(n\). The representation remains
exact, although individual current factors can then have either sign.

\subsection{Finite-current tensor network}

The exact current sum in Equation~\eqref{eq:flux} is
\begin{equation*}
Z=
\sum_{\{n_{ij}\}\in\Z^E}
\prod_{(i,j)\in E}I_{n_{ij}}(\beta J_{ij})
\prod_{k\in V}I_{(\divg n)_k}(\beta h_k),
\end{equation*}
and its finite-current approximation, Equation~\eqref{eq:truncatedflux}, is
\begin{equation*}
Z_K=
\sum_{\{n_{ij}\}\in\mathcal N_K^E}
\prod_{(i,j)\in E}I_{n_{ij}}(\beta J_{ij})
\prod_{k\in V}I_{(\divg n)_k}(\beta h_k),
\qquad
\mathcal N_K=\{-K,\ldots,K\},\quad D=2K+1.
\end{equation*}
Every edge current now labels a bond of dimension \(D\), so evaluating \(Z_K\) is a finite
tensor-network contraction. We next derive the contractions for the partition function and
one-site observables; the resulting scores are mapped to capital weights in
Section~\ref{sec:portfolio-map}.

\section{Tensor-network contraction and observables}
\label{sec:tensor-contraction}
\subsection{Bottom-up contraction of the partition function}
Root the loop-free graph at \(r\) and orient each edge from parent to child. Write \(p(i)\)
for the parent of node \(i\), \(\mathcal C(i)\) for its children, and \(n_{pi}\) and
\(n_{ic}\) for the corresponding currents. Assigning each edge factor to its parent gives the
local tensor
\begin{equation}
\mathcal T_i\!\left(n_{pi},\{n_{ic}\}_{c\in\mathcal C(i)}\right)
=
I_{\sum_c n_{ic}-n_{pi}}(\beta h_i)
\prod_{c\in\mathcal C(i)}I_{n_{ic}}(\beta J_{ic}),
\label{eq:treetensor}
\end{equation}
with \(n_{p(r)r}=0\). A subtree is compressed into the upward message
\begin{equation}
\mu_{i\to p(i)}(n_{pi})
=
\sum_{\{n_{ic}\}}
\mathcal T_i\!\left(n_{pi},\{n_{ic}\}\right)
\prod_{c\in\mathcal C(i)}\mu_{c\to i}(n_{ic}).
\label{eq:upmessage}
\end{equation}
Starting at the leaves, each message contracts its subtree into the index connecting it to the
parent. Once all messages reach the root, the remaining contraction gives
\begin{equation}
Z_K=
\sum_{\{n_{rc}\}}
\mathcal T_r\!\left(0,\{n_{rc}\}\right)
\prod_{c\in\mathcal C(r)}\mu_{c\to r}(n_{rc}).
\label{eq:rootZ}
\end{equation}

\begin{figure}[htbp]
\centering
\begin{subfigure}[b]{0.46\textwidth}
\centering
\begin{tikzpicture}[
  x=0.85cm,y=0.75cm,
  tn/.style={draw,circle,fill=MidnightBlue!10,minimum size=5.5mm,inner sep=0pt},
  msg/.style={->,>=stealth,line width=0.8pt,MidnightBlue}]
  \node[tn] (r) at (0,2.8) {\(r\)};
  \node[tn] (a) at (-1.2,1.55) {};
  \node[tn] (b) at (1.2,1.55) {};
  \node[tn] (c) at (-1.8,0.3) {};
  \node[tn] (d) at (-0.6,0.3) {};
  \node[tn] (e) at (0.6,0.3) {};
  \node[tn] (f) at (1.8,0.3) {};
  \draw[msg] (c)--(a); \draw[msg] (d)--(a);
  \draw[msg] (e)--(b); \draw[msg] (f)--(b);
  \draw[msg] (a)--(r); \draw[msg] (b)--(r);
  \node[MidnightBlue,font=\scriptsize] at (0,-0.25) {upward subtree messages \(\mu\)};
\end{tikzpicture}
\caption{Bottom-up sweep for \(Z_K\)}
\end{subfigure}\hfill
\begin{subfigure}[b]{0.50\textwidth}
\centering
\begin{tikzpicture}[
  x=0.85cm,y=0.75cm,
  tn/.style={draw,circle,fill=MidnightBlue!10,minimum size=5.5mm,inner sep=0pt},
  target/.style={draw,circle,fill=Goldenrod!35,minimum size=6.3mm,inner sep=0pt},
  up/.style={->,>=stealth,line width=0.65pt,MidnightBlue},
  down/.style={->,>=stealth,line width=0.8pt,BrickRed}]
  \node[tn] (r2) at (0,2.8) {\(r\)};
  \node[target] (a2) at (-1.2,1.55) {\(i\)};
  \node[tn] (b2) at (1.2,1.55) {};
  \node[tn] (c2) at (-1.8,0.3) {};
  \node[tn] (d2) at (-0.6,0.3) {};
  \node[tn] (e2) at (0.6,0.3) {};
  \node[tn] (f2) at (1.8,0.3) {};
  \draw[up] (c2)--(a2); \draw[up] (d2)--(a2);
  \draw[up] (e2)--(b2); \draw[up] (f2)--(b2);
  \draw[down] (r2)--(a2); \draw[down] (r2)--(b2);
  \draw[down] (b2)--(e2); \draw[down] (b2)--(f2);
  \node[font=\scriptsize] at (0,-0.25) {
    \textcolor{MidnightBlue}{upward \(\mu\)} \(+\)
    \textcolor{BrickRed}{downward \(\nu\)}};
\end{tikzpicture}
\caption{Top-down environments for node \(i\)}
\end{subfigure}
\caption{Two-pass contraction on a rooted tree. The upward pass produces the partition function;
the downward pass combines the complementary part of the tree with each node's child messages to
obtain its marginal. The empirical path is the degree-two special case.}
\label{fig:treepasses}
\end{figure}
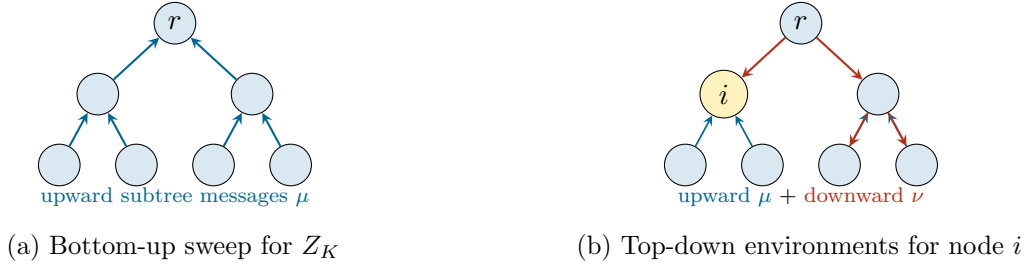

\paragraph{Path specialisation.}
For the Ward order \(\pi(1),\ldots,\pi(N)\), orient the path from left to right. Denote the
edge currents by \(n_\ell\) and impose \(n_0=n_N=0\). Equation~\eqref{eq:rootZ} becomes
\begin{equation}
Z_K=
\sum_{n_1,\ldots,n_{N-1}\in\mathcal N_K}
\prod_{\ell=1}^{N-1}I_{n_\ell}(\beta J_{\pi(\ell),\pi(\ell+1)})
\prod_{\ell=1}^{N}I_{n_\ell-n_{\ell-1}}(\beta h_{\pi(\ell)}).
\label{eq:pathZ}
\end{equation}
For \(a,b\in\mathcal N_K\), define
\begin{align}
M_\ell(a,b)&=I_{b-a}(\beta h_{\pi(\ell)}),&
W_\ell(b,b)&=I_b(\beta J_{\pi(\ell),\pi(\ell+1)}).
\label{eq:transferfactors}
\end{align}
If \(e_0\) is the basis vector for zero current, then
\begin{equation}
Z_K=e_0^\top M_1W_1M_2W_2\cdots M_{N-1}W_{N-1}M_Ne_0.
\label{eq:transferZ}
\end{equation}
On the path, the upward sweep of Figure~\ref{fig:treepasses}(a) becomes a left-to-right recursion,
and the downward environment sweep becomes a right-to-left recursion:
\begin{align}
L_1^\top&=e_0^\top,&
L_{\ell+1}^\top&=L_\ell^\top M_\ell W_\ell,\label{eq:leftenv}\\
R_N&=e_0,&
R_{\ell-1}&=W_{\ell-1}M_\ell R_\ell.\label{eq:rightenv}
\end{align}
At each site, \(L_\ell^\top M_\ell R_\ell=Z_K\). To stabilise the calculation, each updated
message is divided by its largest absolute entry. Accumulated logarithmic scales recover
\(\log Z_K\) and cancel in observable ratios. Dense site updates cost \(O(D^2)\), diagonal
bond updates cost \(O(D)\), and storing all environments requires \(O(ND)\) memory.

\subsection{Top-down environments and one-site marginals}
To evaluate a local marginal, the upward subtree message must be combined with the rest of the
graph. The downward message from parent \(p\) to child \(i\) contracts the parent tensor,
its incoming downward environment, and all sibling subtrees:
\begin{equation}
\begin{aligned}
\nu_{p\to i}(n_{pi})
={}&
\sum_{n_{gp}}
\sum_{\{n_{pj}:j\in\mathcal C(p)\setminus\{i\}\}}
\nu_{g\to p}(n_{gp})\\
&\times
\mathcal T_p\!\left(n_{gp},n_{pi},\{n_{pj}\}_{j\ne i}\right)
\prod_{j\in\mathcal C(p)\setminus\{i\}}\mu_{j\to p}(n_{pj}),
\end{aligned}
\label{eq:downmessage}
\end{equation}
The root boundary is \(\nu_{\varnothing\to r}(0)=1\). This recursion gives the red downward
sweep in Figure~\ref{fig:treepasses}(b). Together with the upward messages from all children,
\(\nu_{p\to i}\) supplies the complete environment of node \(i\) and hence its marginal,
\begin{equation}
p_i(\theta_i)
=
\frac{1}{2\pi Z_K}
\prod_{k\ne i}
\int_0^{2\pi}\frac{d\theta_k}{2\pi}
\exp[-\beta H(\bm\theta)].
\label{eq:marginal}
\end{equation}

Inserting \(e^{iq\theta_{\pi(\ell)}}\) in the current representation changes only the tensor
at the target site. On the path, the shifted tensor is
\begin{equation}
M_\ell^{(q)}(a,b)=I_{b-a+q}(\beta h_{\pi(\ell)}),
\qquad q\in\mathbb Z,
\label{eq:shiftedM}
\end{equation}
so the required Fourier moment of the marginal is
\begin{equation}
\expect{e^{iq\theta_{\pi(\ell)}}}_K
=
\int_0^{2\pi}p_{\pi(\ell)}(\theta)e^{iq\theta}\,d\theta
=
\frac{L_\ell^\top M_\ell^{(q)}R_\ell}
{L_\ell^\top M_\ell R_\ell}
=
\frac{Z_{\pi(\ell)}^{(q)}}{Z_K}.
\label{eq:chargemoment}
\end{equation}
Setting \(q=0\) returns one and verifies marginal normalisation.

\subsection{From marginals to equilibrium cosine scores}
The first cosine moment of the marginal gives the allocation score. Using
\(\cos\theta=(e^{i\theta}+e^{-i\theta})/2\), we obtain
\begin{align}
m_i
&=\int_0^{2\pi}p_i(\theta)\cos\theta\,d\theta \notag\\
&=\frac{1}{2}\left[
\int_0^{2\pi}p_i(\theta)e^{i\theta}\,d\theta+
\int_0^{2\pi}p_i(\theta)e^{-i\theta}\,d\theta
\right]\notag\\
&=\frac{1}{2}\left(
\expect{e^{i\theta_i}}_K+\expect{e^{-i\theta_i}}_K
\right)
=
\frac{Z_i^{(+1)}+Z_i^{(-1)}}{2Z_K}.
\label{eq:cos}
\end{align}
The charged partition functions \(Z_i^{(+1)}\) and \(Z_i^{(-1)}\) are obtained by inserting
\(e^{+i\theta_i}\) and \(e^{-i\theta_i}\) at asset \(i\). For \(i=\pi(\ell)\), the
stored environments give \(Z_i^{(\pm1)}=L_\ell^\top M_\ell^{(\pm1)}R_\ell\). Only the
local tensor is replaced by its shifted form, Equation~\eqref{eq:shiftedM}; every other tensor is
reused. For real model parameters the two inserted moments are conjugates, and \(m_i\) is real.
The response identity \eqref{eq:response}, independent-site limit \eqref{eq:fieldonly}, and
\(q=0\) normalisation test provide independent checks of the implementation.

\subsection{Computational complexity}
A complete graph of \(N\) assets contains \(\binom{N}{2}\) edges. Direct quadrature of
Equation~\eqref{eq:partition} is impractical beyond very small \(N\), and exact tensor-network
contraction is exponential in graph treewidth. The flux transformation changes the representation
without removing this dense-graph difficulty. On the path of Section~\ref{sec:geometry}, however,
there are only \(N-1\) edges. With \(D=2K+1\), two environment sweeps and all one-site
insertions require \(O(ND^2)\) time and \(O(ND)\) memory. These costs apply to the sparsified
model. Replacing the dense graph changes the model, with no known error bound relative to the
complete interaction graph.

\section{From equilibrium scores to portfolio weights}
\label{sec:portfolio-map}

\subsection{Softmax allocation and long-only feasibility}
The magnetisation vector \(m=(m_1,\ldots,m_N)\) is mapped to portfolio weights by
\begin{equation}
    w_i
    =
    \frac{\exp(\gamma m_i)}
    {\sum_{j=1}^N \exp(\gamma m_j)}.
\label{eq:softmax}
\end{equation}
The parameter \(\gamma\ge0\) sets the allocation's sensitivity to differences in equilibrium
scores, after the Gibbs distribution has been fixed. We subtract \(\max_j(\gamma m_j)\) from
each exponent before normalisation to avoid overflow; this leaves the weights unchanged.

Because exponentials are strictly positive,
\begin{equation}
    w_i>0,
    \qquad
    \sum_{i=1}^N w_i
    =
    \frac{\sum_i e^{\gamma m_i}}{\sum_j e^{\gamma m_j}}
    =1.
\label{eq:longonly}
\end{equation}
The weights are thus long-only and fully invested, without an additional projection or
constrained optimisation. Their relative sizes satisfy
\begin{equation}
    \frac{w_i}{w_j}
    =
    \exp\!\left[\gamma(m_i-m_j)\right].
\label{eq:weightratio}
\end{equation}
The tensor contraction determines \(m_i-m_j\); the softmax controls how strongly these
differences affect the capital allocation.

\subsection{Interpretation of \texorpdfstring{\(\beta\)}{beta} and
\texorpdfstring{\(\gamma\)}{gamma}}
The parameters enter at different stages. Within the Gibbs model, \(\beta\) multiplies both
couplings and fields: the Bessel factors and marginals depend on \(\beta J_{ij}\) and
\(\beta h_i\). As \(\beta\to0\), the angles become uniformly distributed, \(m_i\to0\),
and any finite \(\gamma\) gives an allocation approaching equal weight. At intermediate
\(\beta\), fields and interactions produce cross-sectional score differences. At large
\(\beta\), low-energy alignment dominates. Such alignment can compress magnetisation
differences, so increasing inverse temperature need not increase portfolio concentration.

After inference, \(\gamma\) changes the allocation without changing \(Z_K\), the marginals,
or \(m_i\). At \(\gamma=0\), Equation~\eqref{eq:softmax} gives \(w_i=1/N\). Larger
\(\gamma\) amplifies the score differences through Equation~\eqref{eq:weightratio}; as
\(\gamma\to\infty\), capital concentrates on the asset or tied assets with maximal \(m_i\).
Thus \(\beta\) determines the scores and \(\gamma\) determines their allocation intensity.
Their effects combine because \(\gamma\) acts on the \(\beta\)-dependent vector
\(m(\beta)\).

\subsection{Effective portfolio breadth}
Portfolio concentration is summarised by the Herfindahl index and its reciprocal effective breadth,
\begin{equation}
    \HHI(\bm w)=\sum_i w_i^2,
    \qquad
    N_{\mathrm{eff}}
    =
    \frac{1}{\sum_i w_i^2}.
\label{eq:neff}
\end{equation}
Effective breadth equals \(N\) for equal weighting and approaches one as the portfolio
concentrates in a single asset. Its units are an equivalent number of equally weighted holdings,
making it a direct diagnostic of allocation concentration.

\section{Numerical results}
\label{sec:results}
\subsection{Data and numerical protocol}
We study fixed large-cap universes in the United States, United Kingdom, Germany, Japan, and
India. Adjusted closing prices over 2016-01-01--2026-01-30 are drawn from the stored
\texttt{yfinance} snapshot. Securities satisfy a 97\% availability rule, and each market is
aligned on a complete-case calendar. The retained asset lists appear in
Appendix~\ref{app:implementation}; Table~\ref{tab:markets} gives the sample sizes. All calculations
use local currencies, without cross-market currency conversion.
\begin{table}[htbp]
\centering
\small
\begin{tabular}{llrrl}
\toprule
Market & Headline index & $N$ & $T$ & Ticker suffix \\
\midrule
US      & S\&P~500 (\texttt{\^{}GSPC})   & 30 & 2533 & none \\
UK      & FTSE~100 (\texttt{\^{}FTSE})   & 30 & 2540 & \texttt{.L} \\
Germany & DAX (\texttt{\^{}GDAXI})       & 29 & 2561 & \texttt{.DE} \\
Japan   & Nikkei~225 (\texttt{\^{}N225}) & 30 & 2483 & \texttt{.T} \\
India   & Nifty~50 (\texttt{\^{}NSEI})   & 29 & 2489 & \texttt{.NS} \\
\bottomrule
\end{tabular}
\caption{Market universes and complete daily return observations. \(N\) is the number of retained
equities and \(T\) the number of observations.}
\label{tab:markets}
\end{table}
The universes are fixed ex-post lists rather than point-in-time constituent histories.
Survivorship and look-ahead bias therefore limit predictive interpretation. Daily log returns
supply the XY fields and correlations; benchmark calculations use daily simple returns and
annualisation by 252.

The method is defined for continuous \(\beta>0\) and \(\gamma\ge0\). To examine its
behaviour, the concentration surfaces in Section~\ref{subsec:concentration-results} sample
\(\beta\in[0.2,16]\) and \(\gamma\in[0,120]\), using the same contraction and softmax
procedures throughout. Six representative pairs illustrate near-equal, intermediate, and
concentrated allocations:
\[
    (1,1),\quad (2,60),\quad (5,10),\quad
    (5,30),\quad (5,120),\quad (10,60).
\]
These manually chosen points illustrate distinct regimes within a continuous parameter space.
They are neither calibrated values nor outputs of an optimisation procedure. Their realised
outcomes are compared on the common 2016--2026 sample.

For portfolio weights \(\bm w\), estimated expected return and volatility are
\[
    \mu_p=\bm w^\top\hat{\bm\mu}_s,
    \qquad
    \sigma_p=\sqrt{\bm w^\top\hat\Sigma_s\bm w}.
\]
The long-only Markowitz frontier is obtained by solving, for a sequence of target returns \(r\),
\begin{equation}
    \sigma_\star^2(r)
    =
    \min_{\bm w}\ \bm w^\top\hat\Sigma_s\bm w
    \quad\text{subject to}\quad
    \bm w^\top\hat{\bm\mu}_s=r,\quad
    \sum_iw_i=1,\quad w_i\ge0.
\label{eq:frontierresults}
\end{equation}
The upper efficient boundary comprises portfolios for which no feasible alternative has higher
estimated return at the same or lower estimated risk. Its maximum zero-rate Sharpe point is the
tangency portfolio. Overlaying the XY allocations locates them within this estimated feasible
set; Equation~\eqref{eq:frontierresults} does not define their optimisation objective.

For daily simple portfolio returns \(R_{p,t}=\bm w^\top\bm R_t\), the reported quantities are
\begin{align}
    \text{annualised return}
    &=252\,\overline{R}_p,\\
    \text{annualised volatility}
    &=\sqrt{252}\,\operatorname{sd}(R_{p,t}),\\
    \text{Sharpe ratio}
    &=\frac{\text{annualised return}}{\text{annualised volatility}},
\end{align}
with the risk-free rate set to zero. Table~\ref{tab:sharpe} compares the highest displayed XY
Sharpe ratio with equal weighting, the headline index, minimum variance, equal-risk contribution,
and the in-sample tangency portfolio.

\subsection{Structure of the empirical interaction network}
\subsubsection{Correlation structure}
The five empirical correlation matrices are shown in Figure~\ref{fig:corr}.

\begin{figure}[p]
\centering
\begin{subfigure}{0.68\textwidth}
  \includegraphics[width=\linewidth]{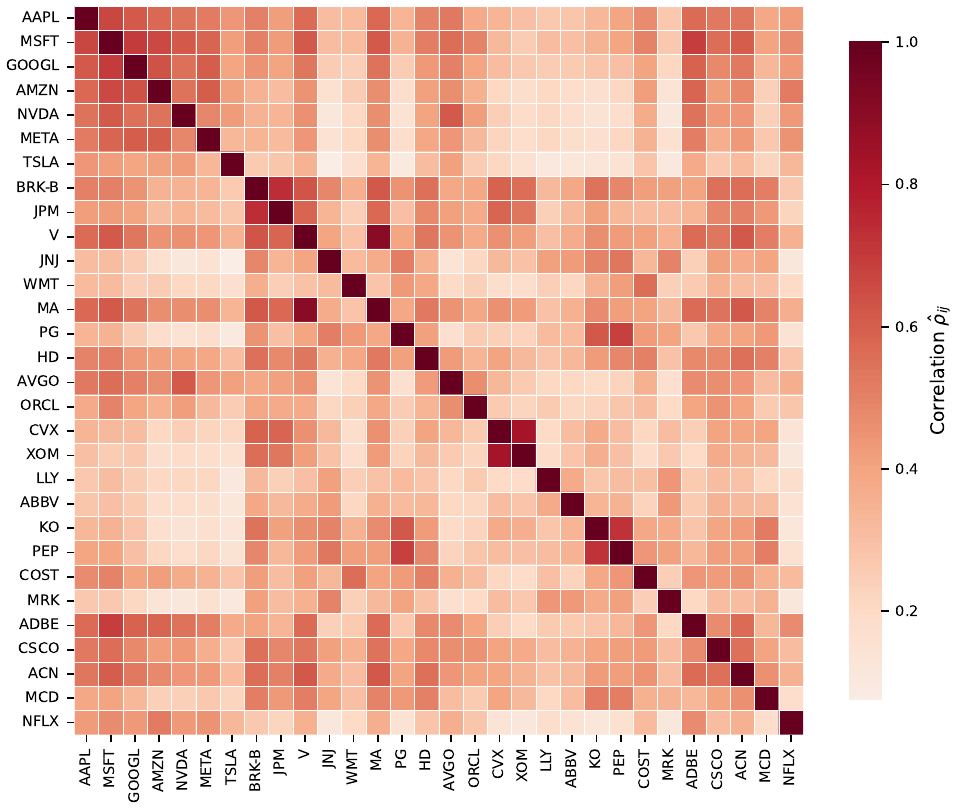}\caption{United States}
\end{subfigure}\par\medskip
\begin{subfigure}{0.68\textwidth}
  \includegraphics[width=\linewidth]{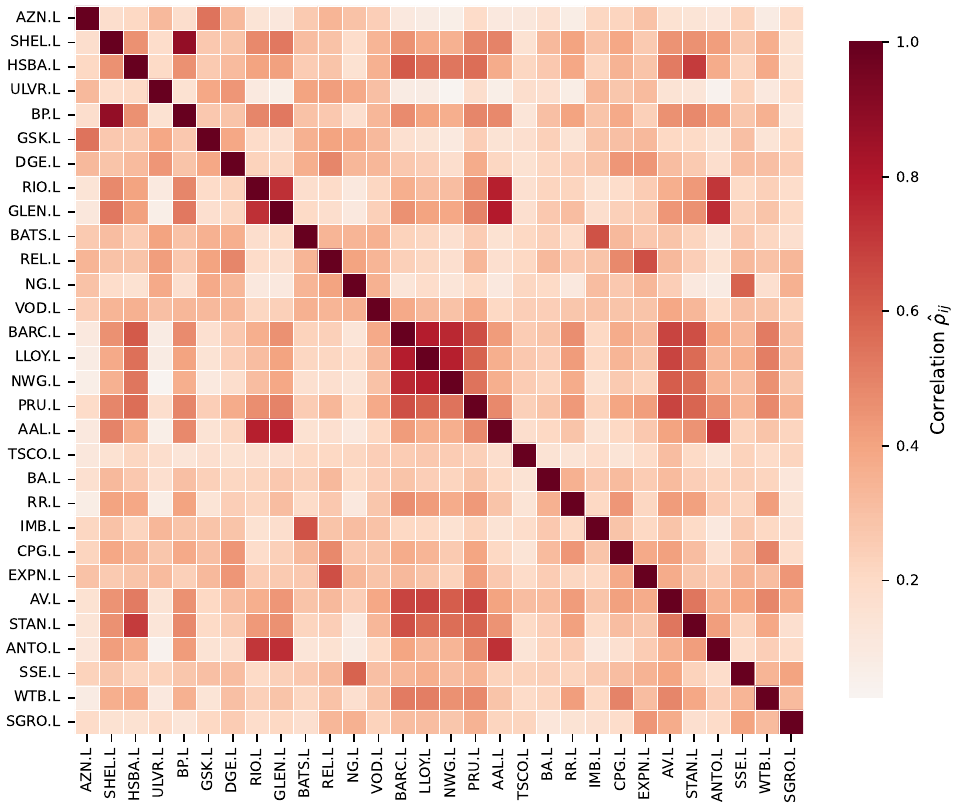}\caption{United Kingdom}
\end{subfigure}
\caption{Sample correlation matrices of daily log returns over 2016--2026. Assets are shown in
the fixed input order; every matrix is estimated from its market's complete-case return panel.}
\label{fig:corr}
\end{figure}

\begin{figure}[p]\ContinuedFloat
\centering
\begin{subfigure}{0.68\textwidth}
  \includegraphics[width=\linewidth]{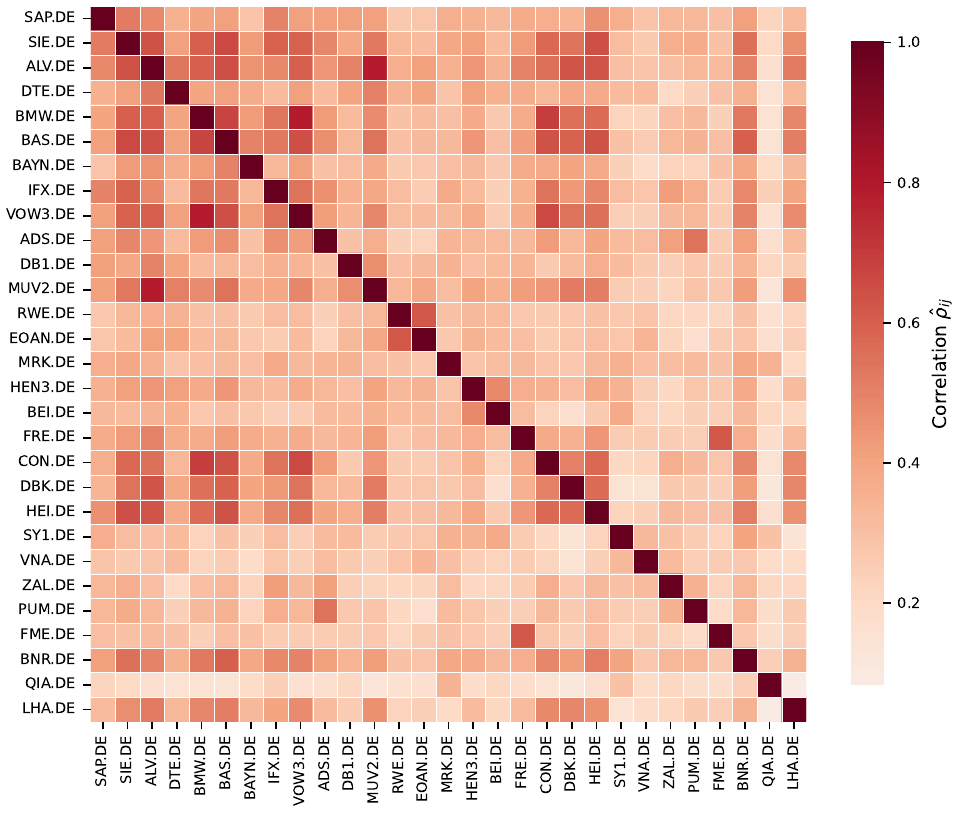}\caption{Germany}
\end{subfigure}\par\medskip
\begin{subfigure}{0.68\textwidth}
  \includegraphics[width=\linewidth]{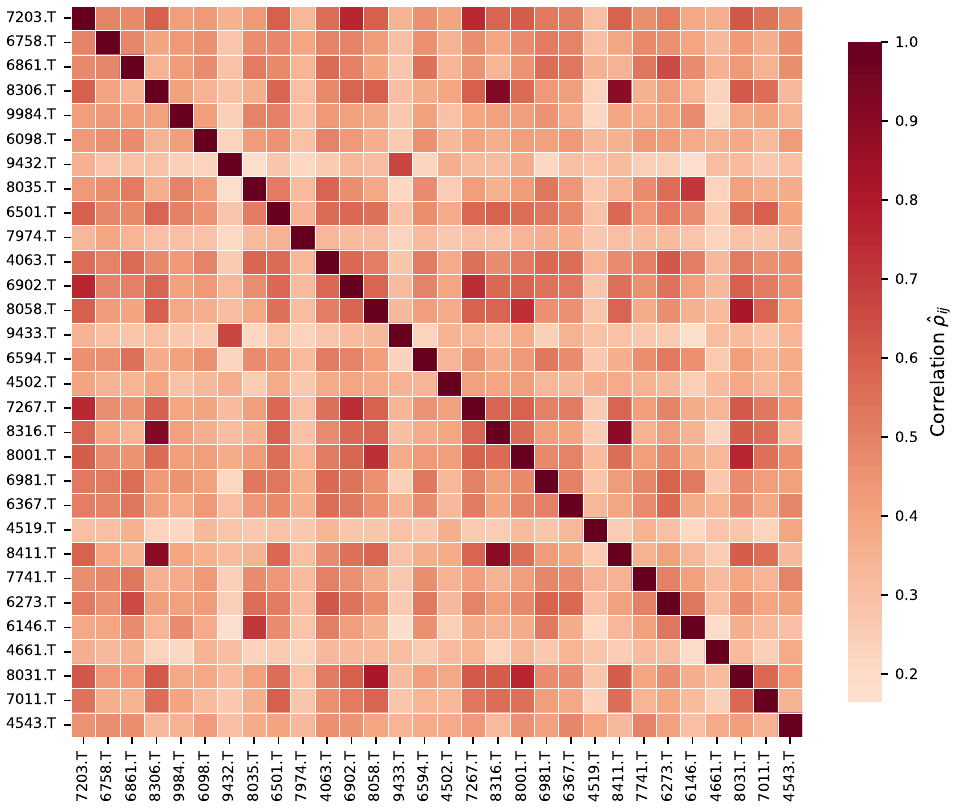}\caption{Japan}
\end{subfigure}
\caption[]{Figure~\ref{fig:corr} continued. Panel conventions follow the first page of the
figure.}
\end{figure}

\begin{figure}[p]\ContinuedFloat
\centering
\begin{subfigure}{0.72\textwidth}
  \includegraphics[width=\linewidth]{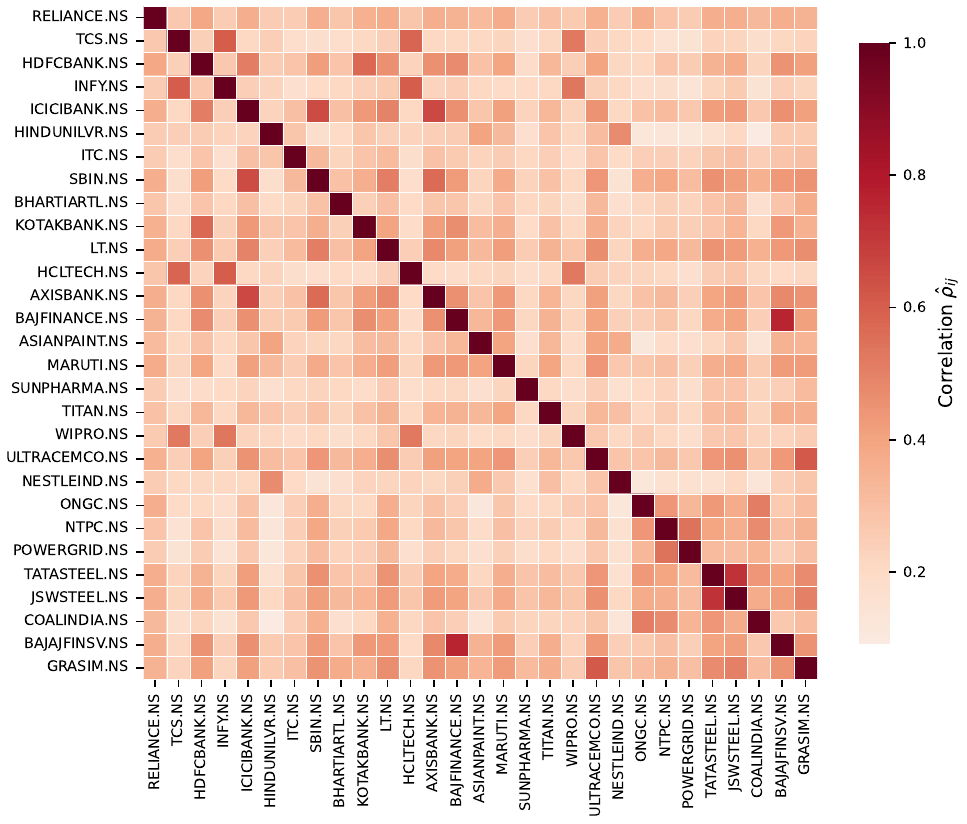}\caption{India}
\end{subfigure}
\caption[]{Figure~\ref{fig:corr} continued. Panel conventions follow the first page of the
figure.}
\end{figure}

The mean off-diagonal correlations are 0.370 for the United States, 0.304 for the United Kingdom,
0.350 for Germany, 0.416 for Japan, and 0.300 for India. In the same order, the strongest pairs
are \texttt{V}--\texttt{MA} (0.899), \texttt{SHEL.L}--\texttt{BP.L} (0.876),
\texttt{ALV.DE}--\texttt{MUV2.DE} (0.784), \texttt{8306.T}--\texttt{8316.T} (0.918), and
\texttt{BAJFINANCE.NS}--\texttt{BAJAJFINSV.NS} (0.760). All 2,117 unique off-diagonal
coefficients are positive in these samples, although the model also permits negative couplings.
Strongly correlated pairs and broader correlation blocks give the heterogeneous structure
encoded by the XY interactions.
\FloatBarrier
\subsubsection{Four-point hyperbolicity of the correlation geometry}
The normalised statistic \(\delta_{\mathrm{worst}}/\operatorname{diam}(d)\) compares the most
discrepant four-asset configuration with the exact tree-metric condition.
\begin{table}[htbp]
\centering
\small
\begin{tabular}{lccccc}
\toprule
Market & $N$ & diameter & $\delta_{\text{worst}}$ &
$\delta_{\text{worst}}/\mathrm{diam}$ & $\delta_{\text{mean}}$ \\
\midrule
US      & 30 & 1.359 & 0.1305 & 0.096 & 0.0248 \\
UK      & 30 & 1.396 & 0.1432 & 0.103 & 0.0266 \\
Germany & 29 & 1.355 & 0.1096 & 0.081 & 0.0223 \\
Japan   & 30 & 1.292 & 0.0868 & 0.067 & 0.0183 \\
India   & 29 & 1.347 & 0.0937 & 0.070 & 0.0175 \\
\bottomrule
\end{tabular}
\caption{Four-point hyperbolicity of \(d_{ij}=\sqrt{2(1-\hat\rho_{ij})}\). The scaled worst-case
quantity is \(\delta_{\mathrm{worst}}/\operatorname{diam}(d)\); the mean is unscaled.}
\label{tab:gromov}
\end{table}

\begin{figure}[p]
\centering
\begin{subfigure}{0.72\textwidth}
  \includegraphics[width=\linewidth]{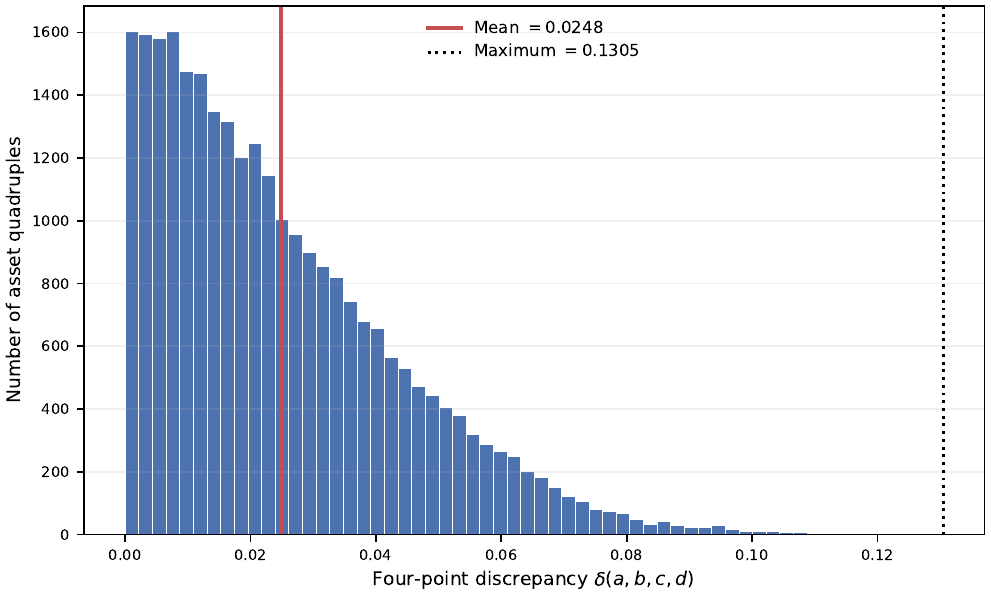}\caption{United States}
\end{subfigure}\par\medskip
\begin{subfigure}{0.72\textwidth}
  \includegraphics[width=\linewidth]{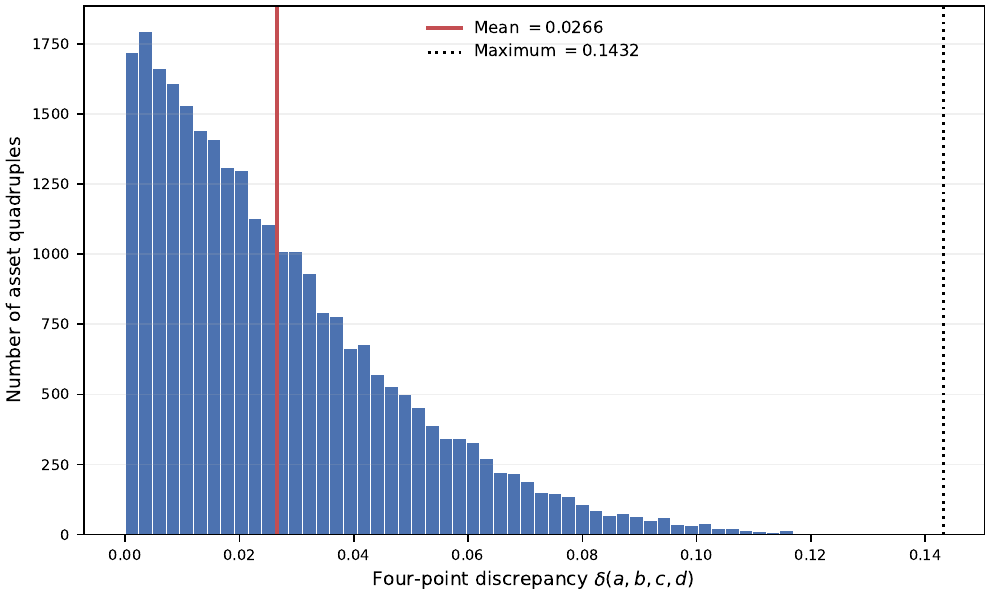}\caption{United Kingdom}
\end{subfigure}
\caption{Distribution of \(\delta(a,b,c,d)\) over all asset quadruples: 27,405 quadruples for
\(N=30\) and 23,751 for \(N=29\). Vertical lines mark the sample mean and maximum.}
\label{fig:gromov}
\end{figure}

\begin{figure}[p]\ContinuedFloat
\centering
\begin{subfigure}{0.72\textwidth}
  \includegraphics[width=\linewidth]{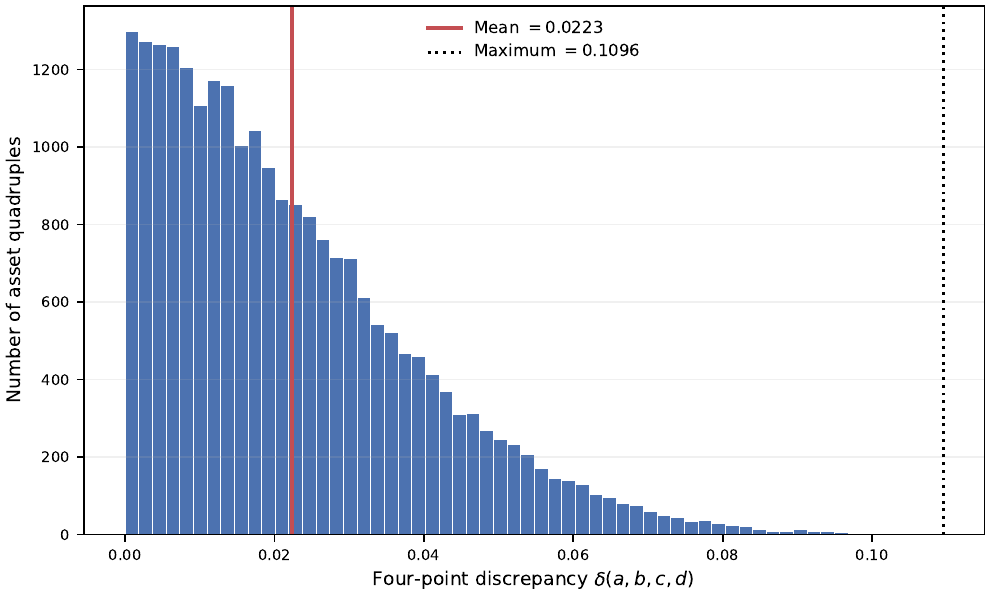}\caption{Germany}
\end{subfigure}\par\medskip
\begin{subfigure}{0.72\textwidth}
  \includegraphics[width=\linewidth]{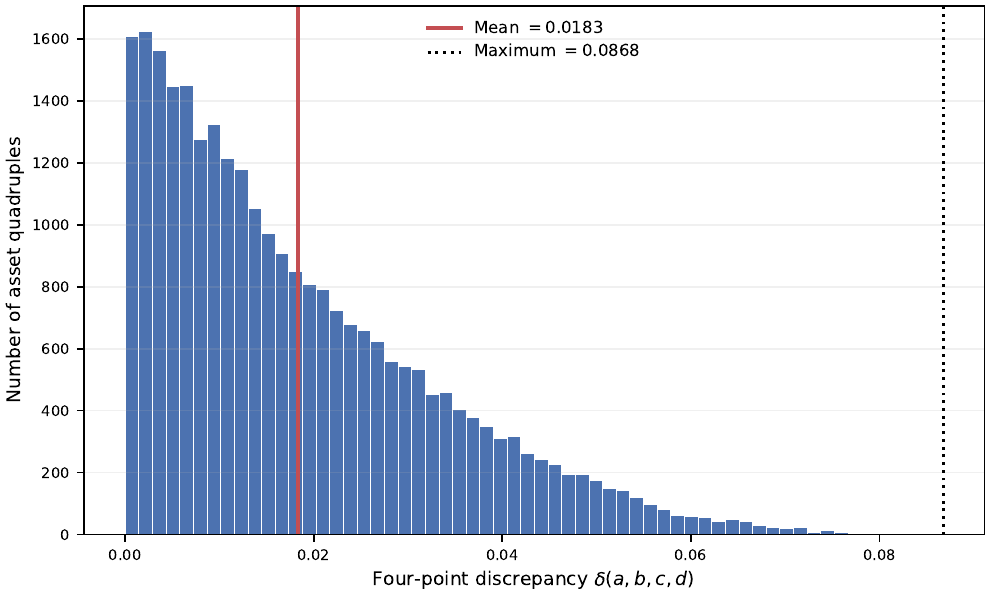}\caption{Japan}
\end{subfigure}
\caption[]{Figure~\ref{fig:gromov} continued. Histograms cover all asset quadruples; vertical
lines mark the mean and maximum discrepancy.}
\end{figure}

\begin{figure}[p]\ContinuedFloat
\centering
\begin{subfigure}{0.78\textwidth}
  \includegraphics[width=\linewidth]{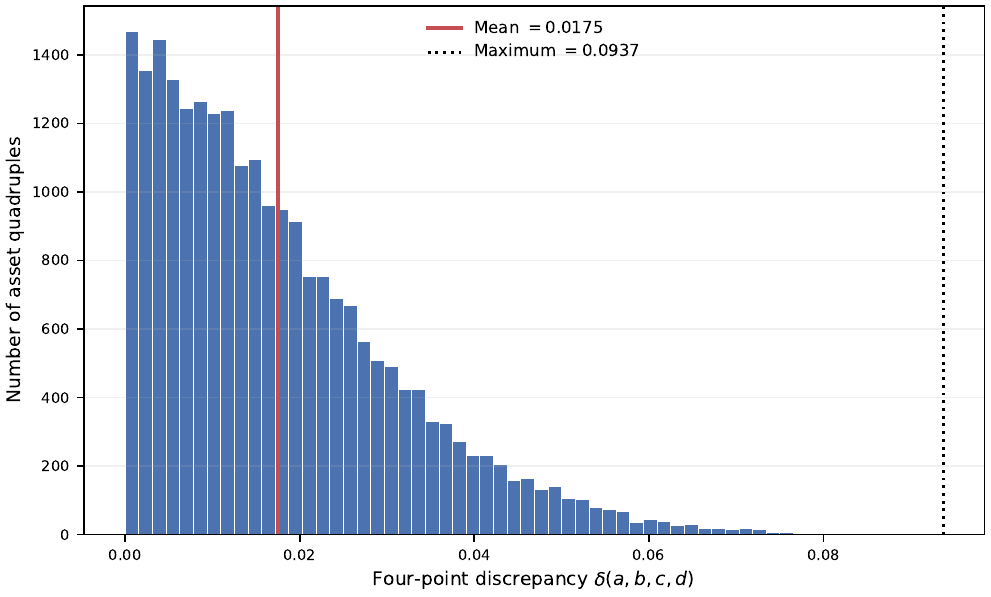}\caption{India}
\end{subfigure}
\caption[]{Figure~\ref{fig:gromov} continued. Panel conventions follow the first page of the
figure.}
\end{figure}

Normalising by the market diameter makes the largest four-point discrepancy comparable across
markets. For \(\delta_{\mathrm{worst}}/\operatorname{diam}(d)\), Japan has the smallest value,
0.067, followed by India (0.070), Germany (0.081), the United States (0.096), and the United
Kingdom (0.103). Japan and India are consequently the most tree-like samples by this diagnostic,
and the United Kingdom is the least. The largest observed discrepancies amount to
6.7--10.3\% of the respective market diameters.

Typical quadruples are closer to tree geometry. Mean discrepancies range from 0.0175 to 0.0266,
and 85.6\%--97.0\% of quadruples satisfy \(\delta<0.05\). None of the distance matrices is
an exact tree metric, but the concentration near zero motivates examining a tree-based surrogate.
There is no universal acceptance threshold. These statistics neither select the Ward path nor
bound the error of replacing the dense XY graph.
\FloatBarrier
\subsubsection{Sparse graph used for tensor-network inference}
After the geometric diagnostic, Ward clustering gives the hierarchy shown in
Figure~\ref{fig:dendro}. Connecting consecutive leaves defines the path used for contraction.
\begin{figure}[p]
\centering
\begin{subfigure}{0.78\textwidth}
  \includegraphics[width=\linewidth]{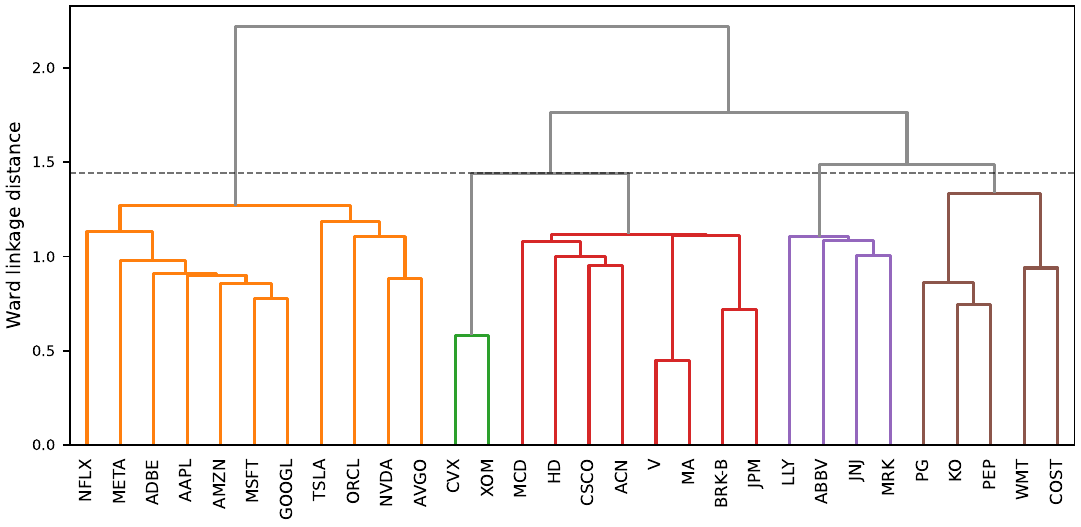}\caption{United States}
\end{subfigure}\par\medskip
\begin{subfigure}{0.78\textwidth}
  \includegraphics[width=\linewidth]{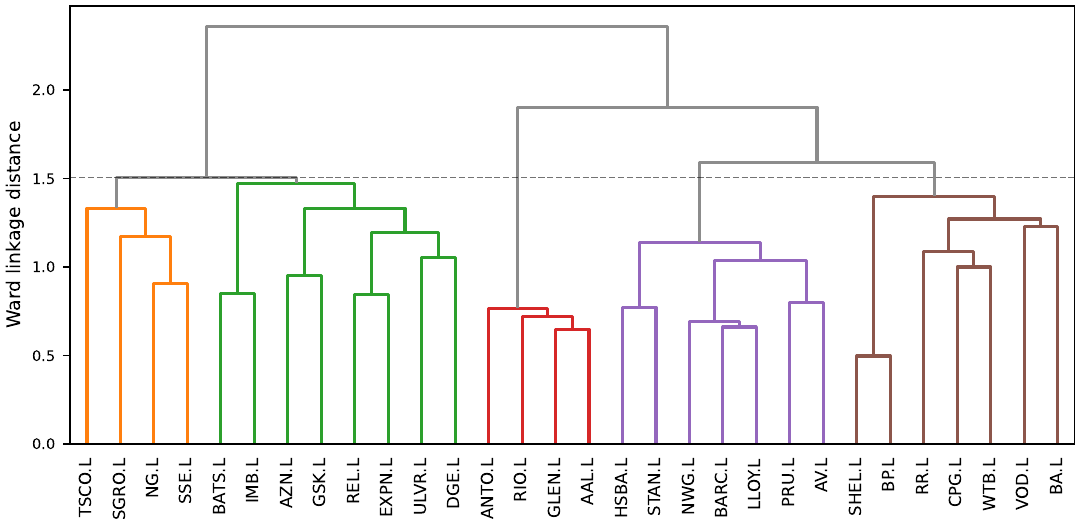}\caption{United Kingdom}
\end{subfigure}
\caption{Ward dendrograms computed from the correlation distance. Dashed lines show the
five-cluster display cut. The interaction path connects successive leaves in the displayed order.}
\label{fig:dendro}
\end{figure}

\begin{figure}[p]\ContinuedFloat
\centering
\begin{subfigure}{0.78\textwidth}
  \includegraphics[width=\linewidth]{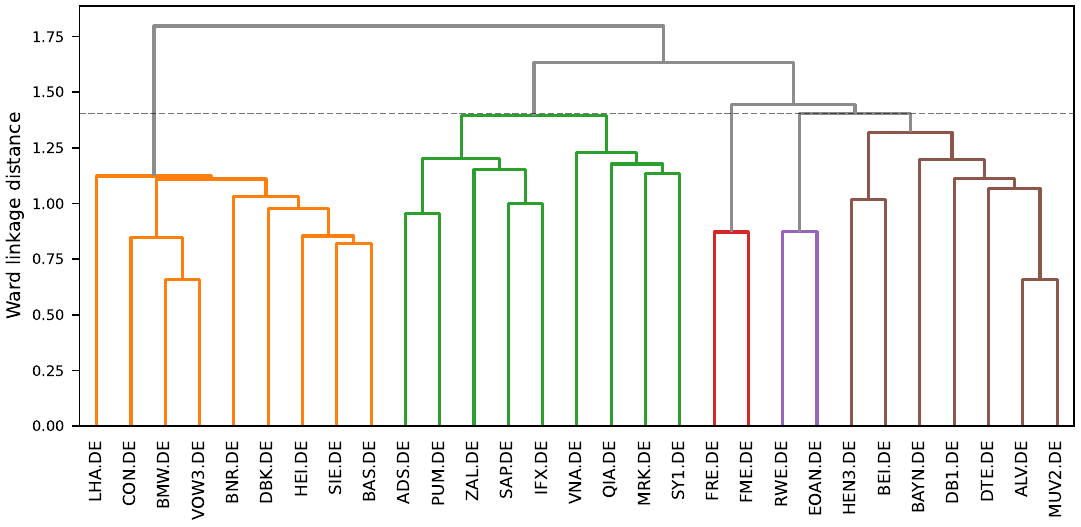}\caption{Germany}
\end{subfigure}\par\medskip
\begin{subfigure}{0.78\textwidth}
  \includegraphics[width=\linewidth]{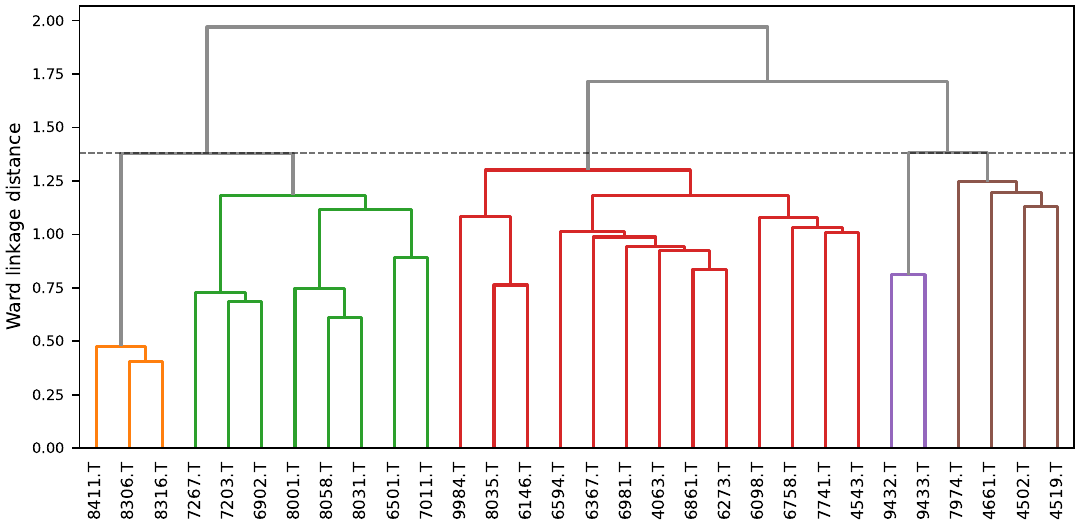}\caption{Japan}
\end{subfigure}
\caption[]{Figure~\ref{fig:dendro} continued. Height is Ward linkage distance; successive leaves
define the interaction path.}
\end{figure}

\begin{figure}[p]\ContinuedFloat
\centering
\begin{subfigure}{0.82\textwidth}
  \includegraphics[width=\linewidth]{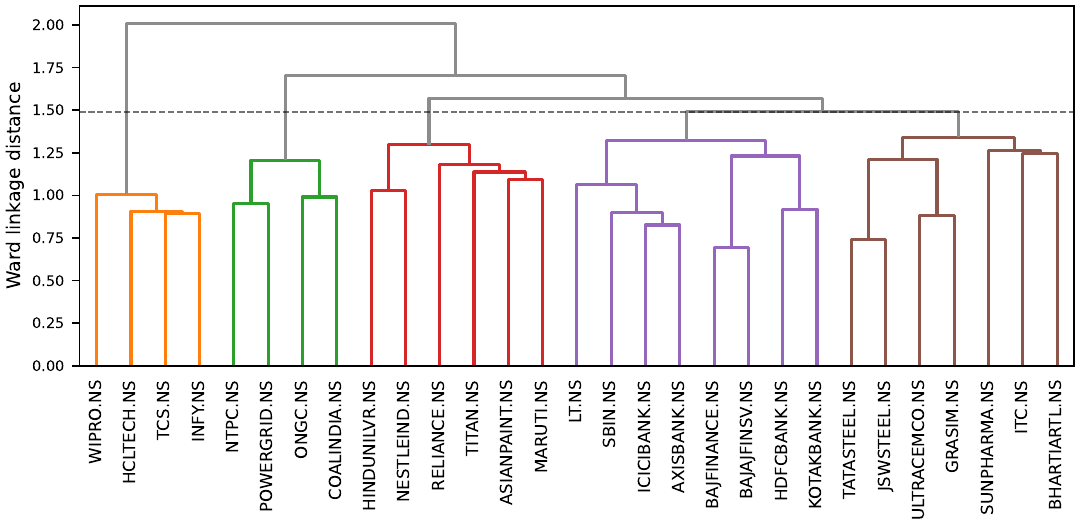}\caption{India}
\end{subfigure}
\caption[]{Figure~\ref{fig:dendro} continued. Panel conventions follow the first page of the
figure.}
\end{figure}

Consecutive-leaf correlations average 0.528, 0.506, 0.463, 0.552, and 0.445 in the US, UK,
Germany, Japan, and India, respectively. Each exceeds its market's complete-graph mean, indicating
that Ward ordering places relatively strong pairs next to one another. Nevertheless, the paths
retain only 8.8--11.1\% of the sum of absolute off-diagonal correlations. The resulting
interaction model is therefore an aggressive sparsification of the dense graph.
\FloatBarrier
\subsection{Numerical validation of the tensor-network calculation}
We check the finite-current contraction in three ways, at the precision relevant to the reported
portfolio weights.

\subsubsection{Current-cutoff convergence}
For each market and \(\beta\in\{0.2,1,2,5,10,16\}\), we repeat the contraction with the
adaptive cutoff \(K\) and the wider window \(K+5\). Across all 30 comparisons,
\(\log Z_K\) is unchanged at double precision. The largest magnetisation difference is
\(\max_i|m_i^{(K+5)}-m_i^{(K)}|=3.4\times10^{-16}\), establishing stability under this
increase of the current window.

\subsubsection{Marginal normalisation}
The \(q=0\) insertion in Equation~\eqref{eq:chargemoment} is evaluated for every asset in the
same 30 market--parameter combinations. All 888 ratios equal one to double precision; the maximum
observed normalisation error is zero at that precision.

\subsubsection{Response-identity check}
For the first, middle, and last assets in each Ward order, we compare the insertion-based
magnetisations with central finite differences of \(\log Z_K\) with respect to \(h_i\).
The 75 checks use \(\beta\in\{0.2,1,5,10,16\}\) and a field step of \(10^{-6}\). The
largest absolute difference between \(m_i\) and
\(\beta^{-1}\partial\log Z_K/\partial h_i\) is \(2.7\times10^{-9}\). Together, these checks
validate contraction of the truncated path model. They do not assess the dense-to-path modelling
error.
\subsection{Dependence on \texorpdfstring{\(\beta\)}{beta} and \texorpdfstring{\(\gamma\)}{gamma}}
\label{subsec:concentration-results}
Figure~\ref{fig:surface} shows how effective breadth \(\Neff=1/\HHI\) depends jointly on
\(\beta\) and \(\gamma\). Breadth near \(N\) indicates almost equal weighting, while
breadth near one indicates extreme concentration. For fixed scores, increasing \(\gamma\)
sharpens the softmax monotonically. The response to \(\beta\) can be non-monotonic because
\(\beta\) changes the scores before allocation.

\begin{figure}[p]
\centering
\begin{subfigure}{0.70\textwidth}
  \includegraphics[width=\linewidth]{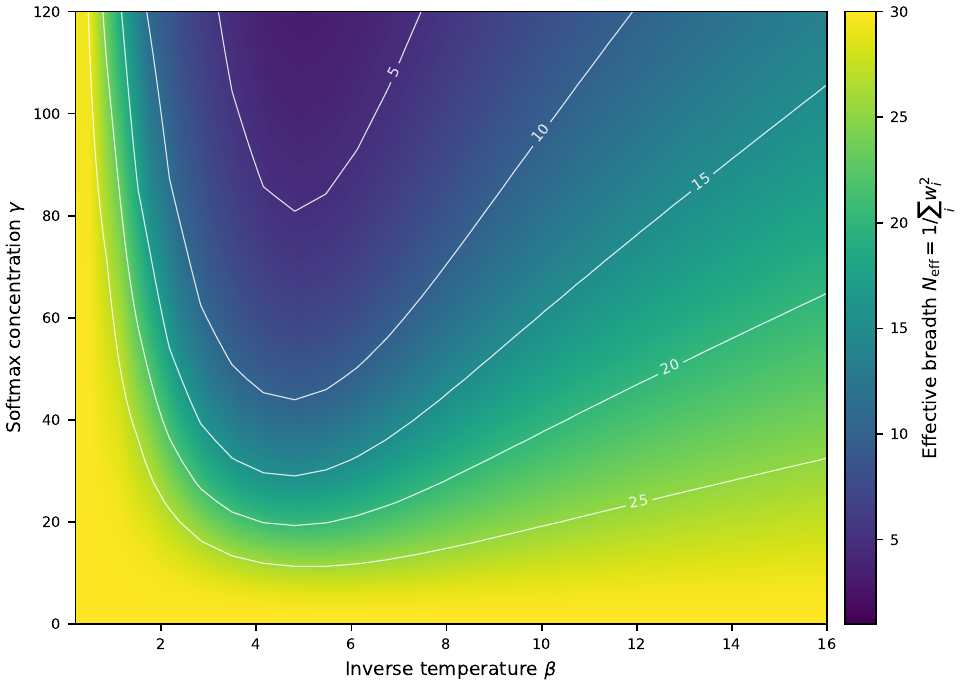}\caption{United States}
\end{subfigure}\par\medskip
\begin{subfigure}{0.70\textwidth}
  \includegraphics[width=\linewidth]{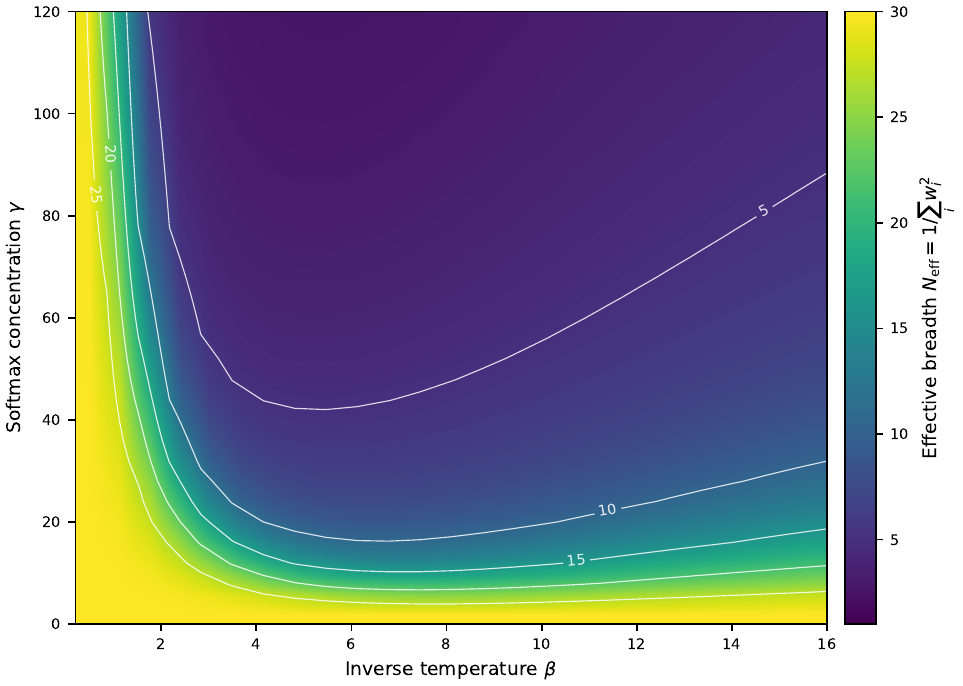}\caption{United Kingdom}
\end{subfigure}
\caption{Effective portfolio breadth \(\Neff=1/\sum_i w_i^2\) over the
\((\beta,\gamma)\) plane. Lighter colours denote broader portfolios and labelled contours give
constant \(\Neff\). These are concentration diagnostics, not realised-performance surfaces.}
\label{fig:surface}
\end{figure}

\begin{figure}[p]\ContinuedFloat
\centering
\begin{subfigure}{0.70\textwidth}
  \includegraphics[width=\linewidth]{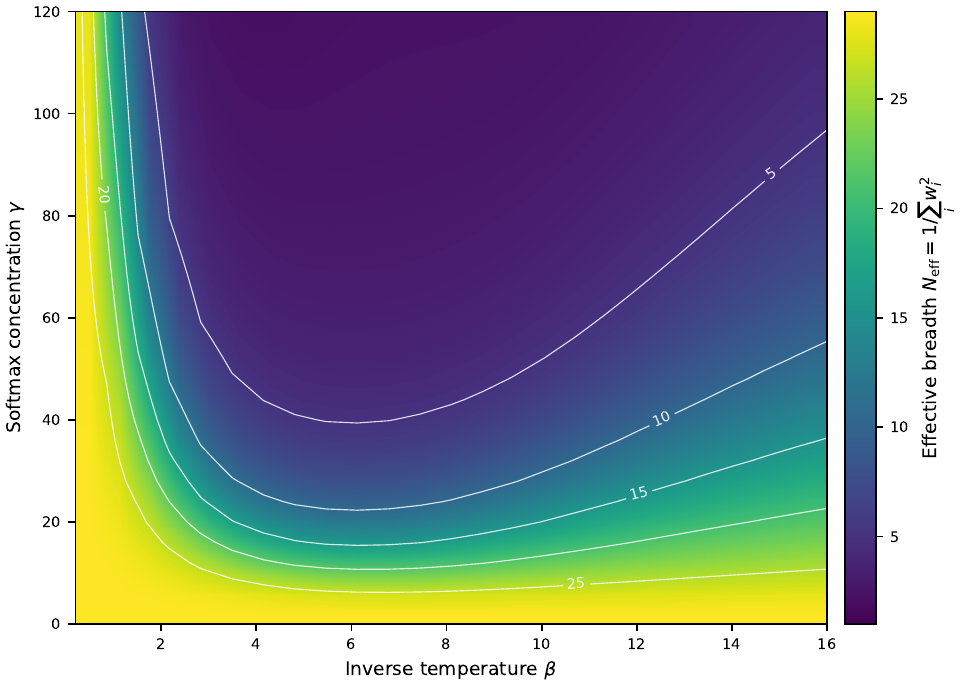}\caption{Germany}
\end{subfigure}\par\medskip
\begin{subfigure}{0.70\textwidth}
  \includegraphics[width=\linewidth]{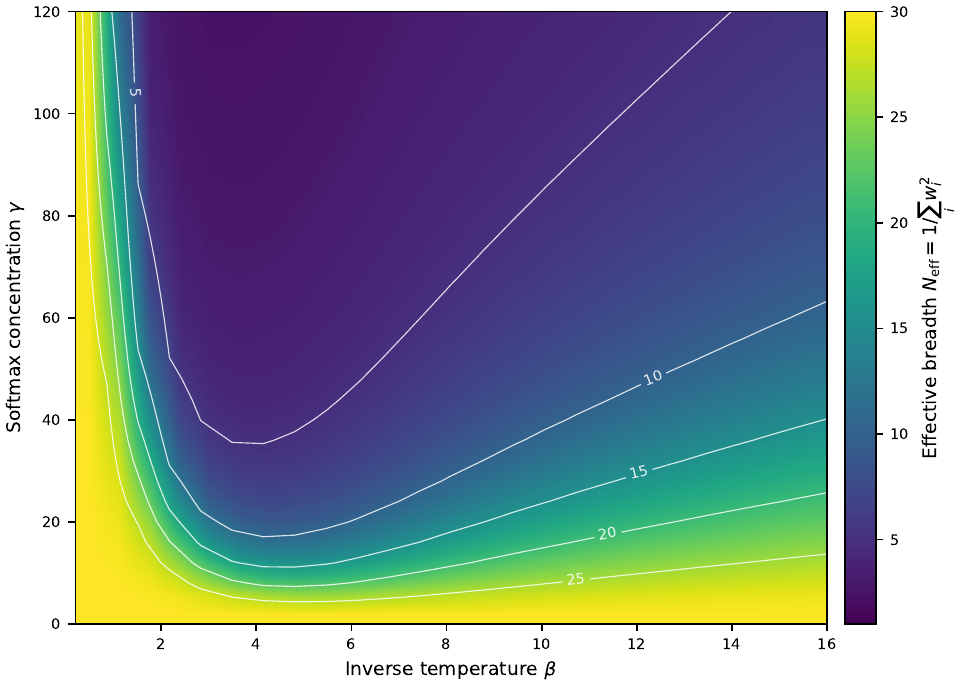}\caption{Japan}
\end{subfigure}
\caption[]{Figure~\ref{fig:surface} continued. Lighter colours denote greater effective breadth;
contours mark constant \(\Neff\).}
\end{figure}

\begin{figure}[p]\ContinuedFloat
\centering
\begin{subfigure}{0.76\textwidth}
  \includegraphics[width=\linewidth]{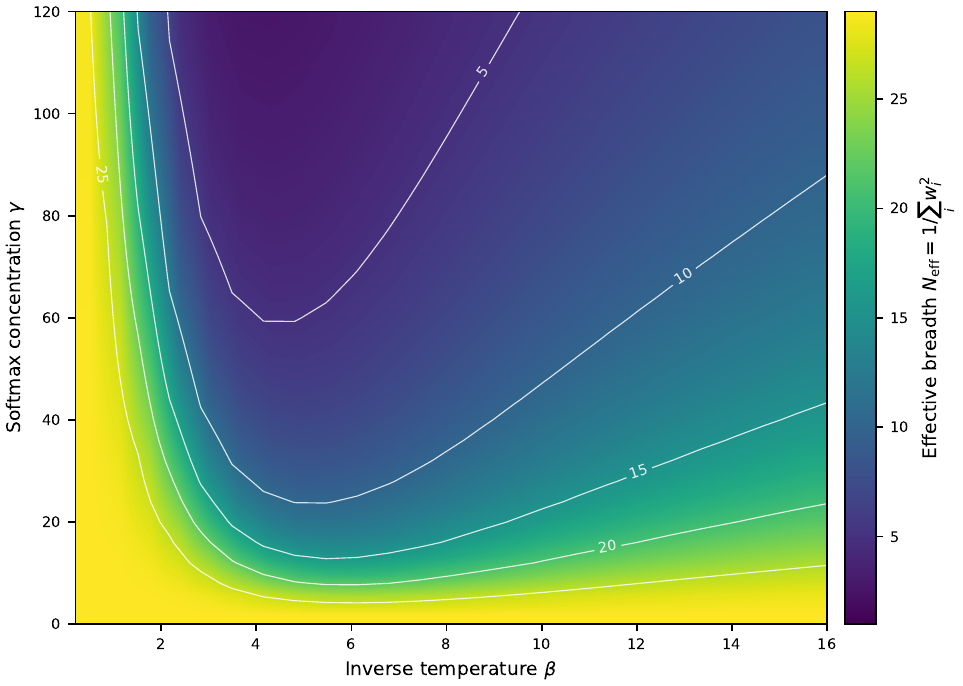}\caption{India}
\end{subfigure}
\caption[]{Figure~\ref{fig:surface} continued. Panel conventions follow the first page of the
figure.}
\end{figure}

At \(\gamma=0\), the softmax is uniform and \(\Neff=N\) in every market, independently of
\(\beta\). The displayed grid reaches its minimum breadth at \(\gamma=120\) and intermediate
inverse temperature: \(\Neff=3.34\) at \(\beta\approx4.81\) in the US, 2.66 at
\(\beta\approx5.47\) in the UK, 2.17 at \(\beta\approx4.15\) in Germany, 2.41 at
\(\beta\approx3.49\) in Japan, and 2.72 at \(\beta\approx4.15\) in India. Beyond these
valleys, increasing \(\beta\) at fixed high \(\gamma\) broadens the allocation. At
\((\beta,\gamma)=(16,120)\), breadth recovers to 13.60, 4.20, 3.88, 5.56, and 8.09,
respectively. Strong low-temperature alignment can therefore reduce cross-sectional score
differences and broaden the portfolio. The valleys describe allocation concentration on a finite
grid; they are not results of parameter optimisation, performance optimisation, or
phase-transition analysis.

\FloatBarrier
\subsection{Representative portfolio allocations}
Figure~\ref{fig:weights} shows the underlying cross-sectional weights for each selected pair and
market. Each panel has one bar per asset, ordered by decreasing allocation.

\begin{figure}[p]
\centering
\makebox[\textwidth][c]{\includegraphics[width=1.10\textwidth]{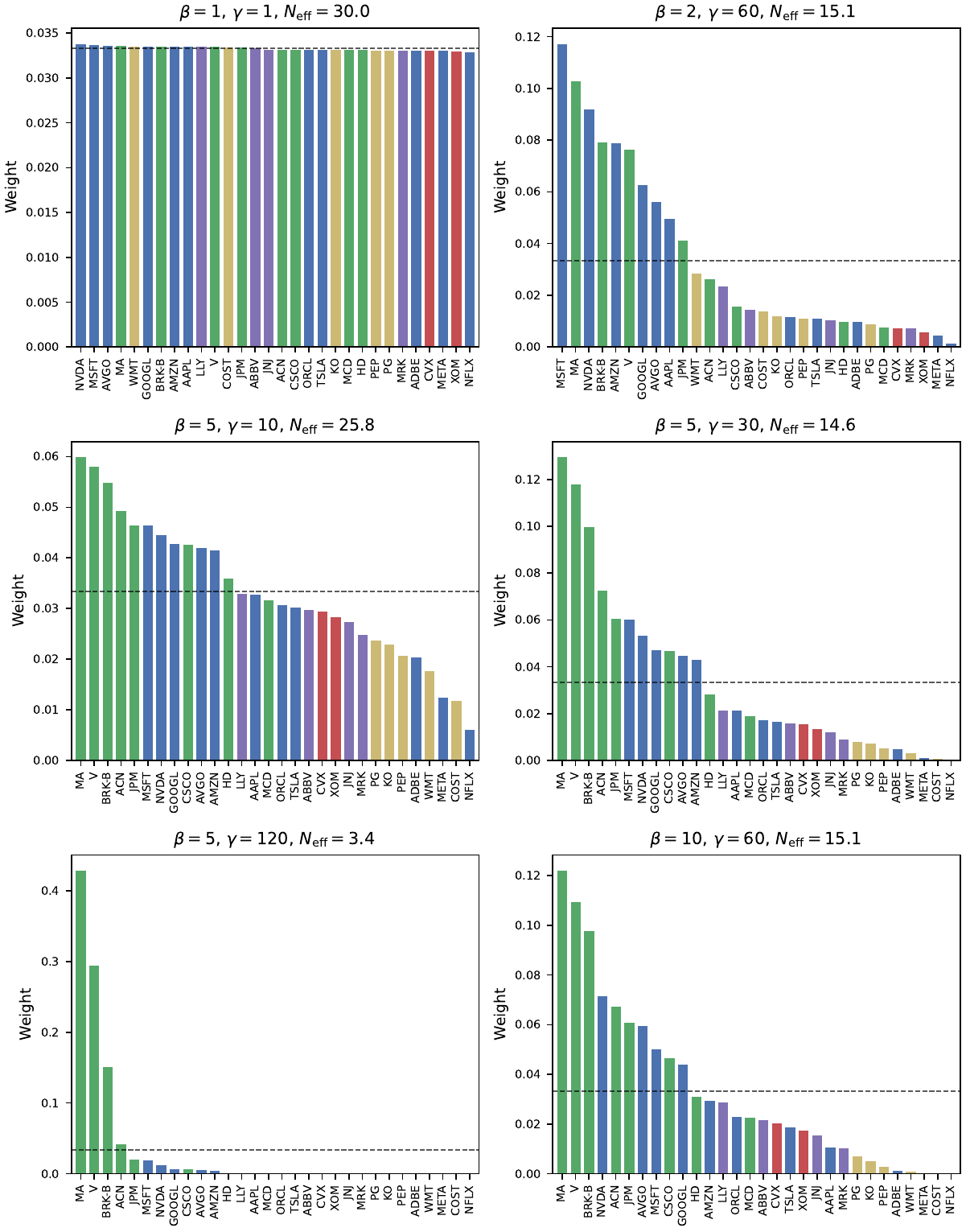}}
\caption{United States portfolio-weight histograms for the six displayed
\((\beta,\gamma)\) pairs. Colours identify Ward clusters, the dashed line is equal weight \(1/N\),
and panel titles report effective breadth.}
\label{fig:weights}
\end{figure}

\begin{figure}[p]\ContinuedFloat
\centering
\makebox[\textwidth][c]{\includegraphics[width=1.10\textwidth]{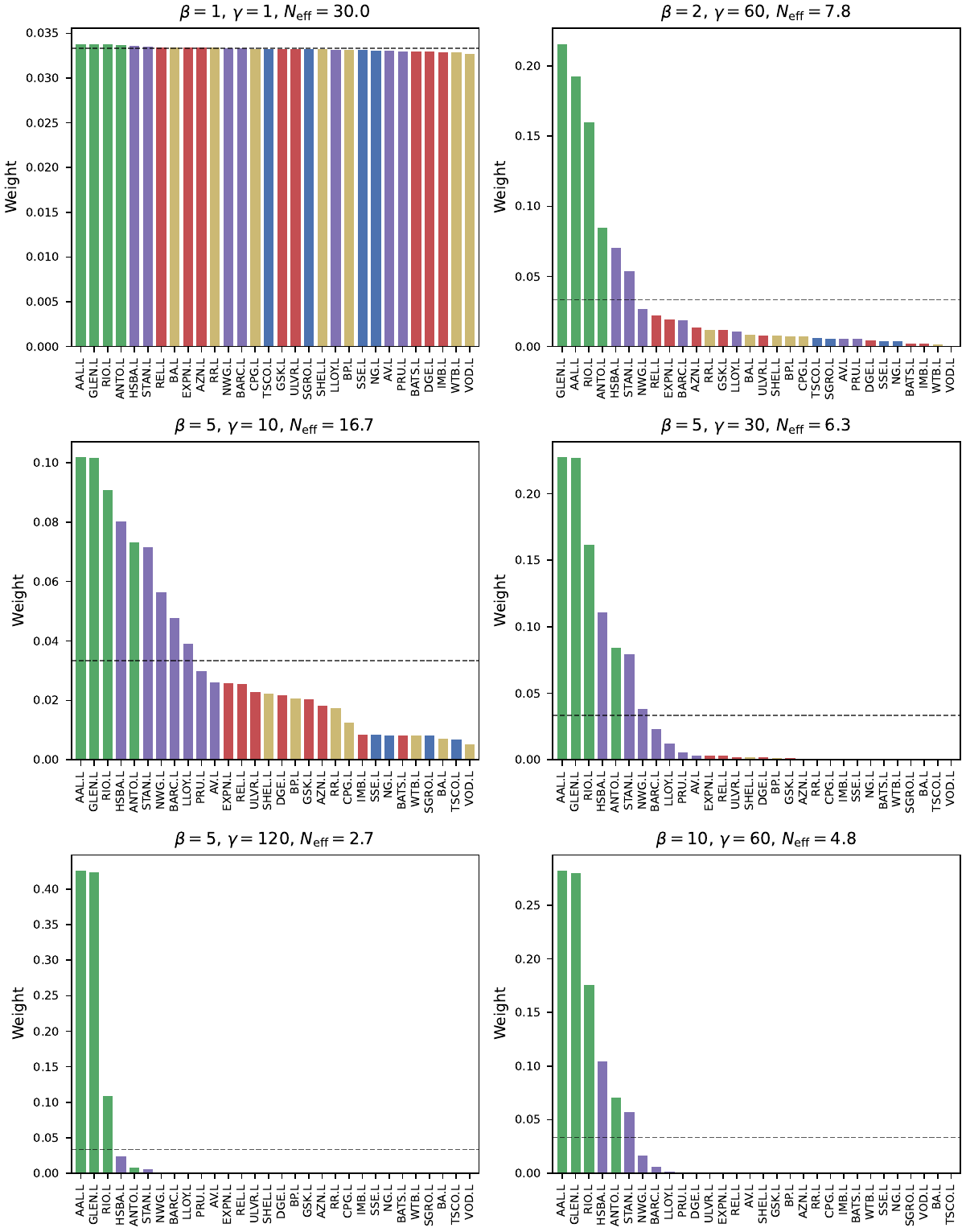}}
\caption[]{Figure~\ref{fig:weights} continued for the United Kingdom. Assets are independently
sorted within each panel.}
\end{figure}

\begin{figure}[p]\ContinuedFloat
\centering
\makebox[\textwidth][c]{\includegraphics[width=1.10\textwidth]{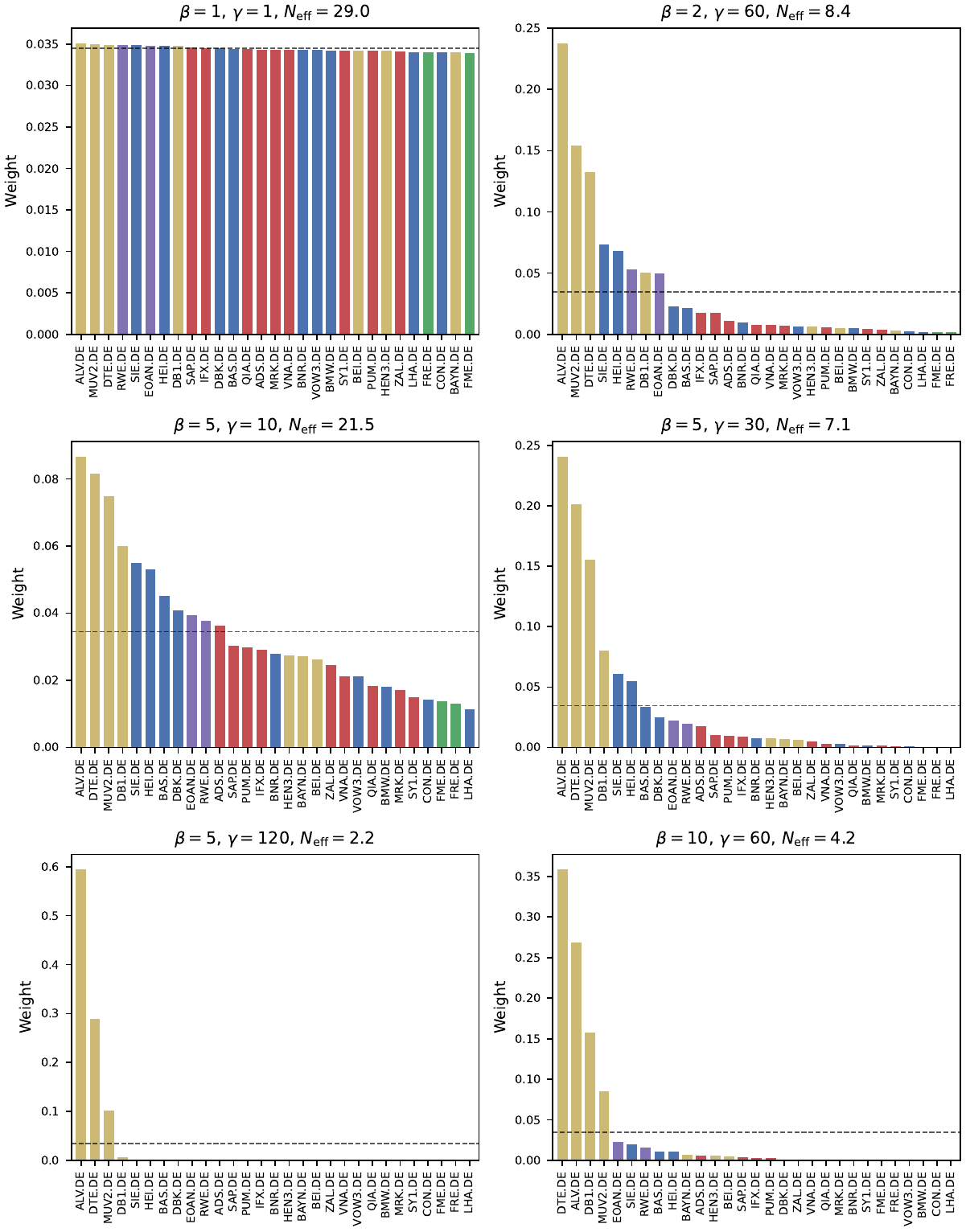}}
\caption[]{Figure~\ref{fig:weights} continued for Germany.}
\end{figure}

\begin{figure}[p]\ContinuedFloat
\centering
\makebox[\textwidth][c]{\includegraphics[width=1.10\textwidth]{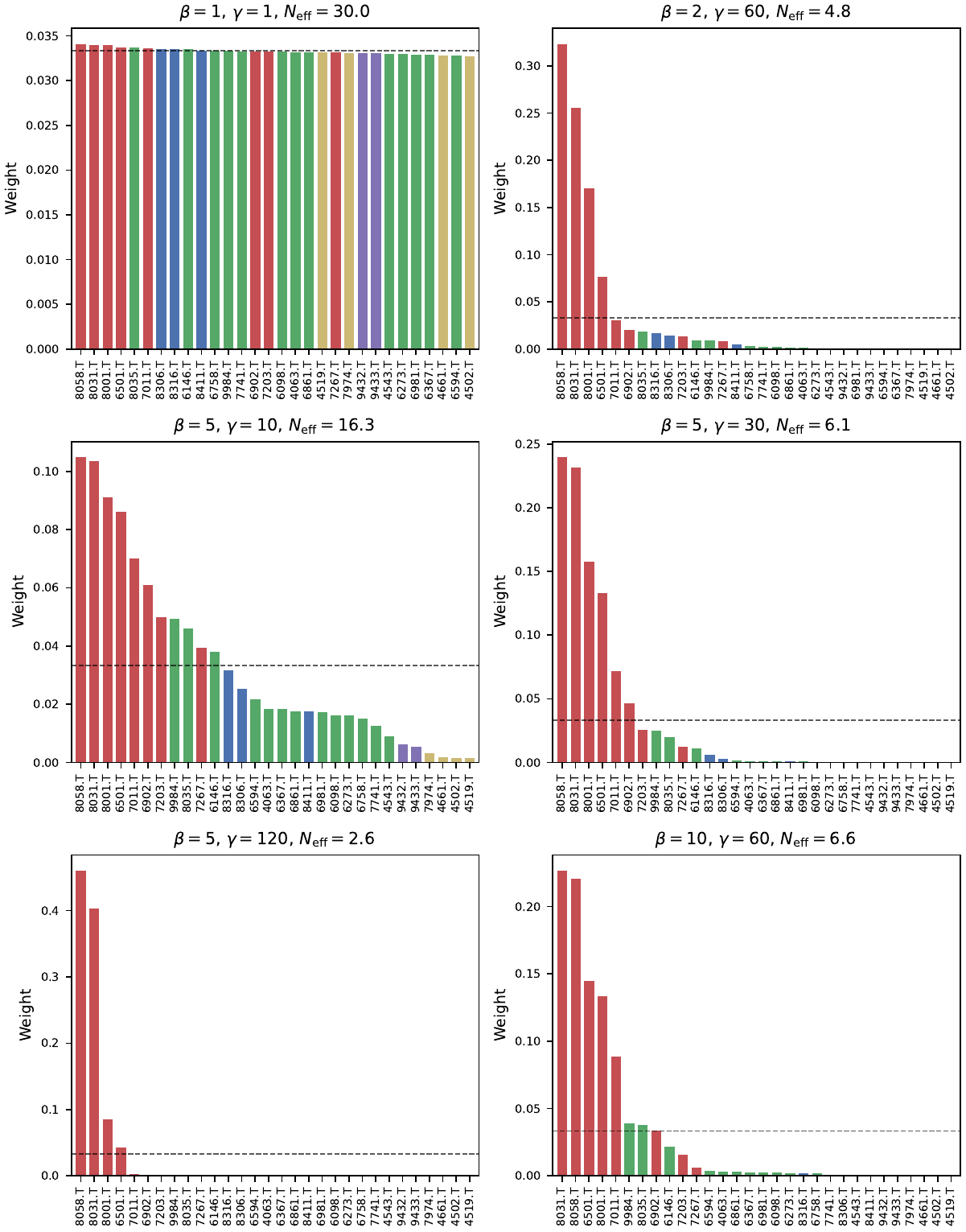}}
\caption[]{Figure~\ref{fig:weights} continued for Japan.}
\end{figure}

\begin{figure}[p]\ContinuedFloat
\centering
\makebox[\textwidth][c]{\includegraphics[width=1.10\textwidth]{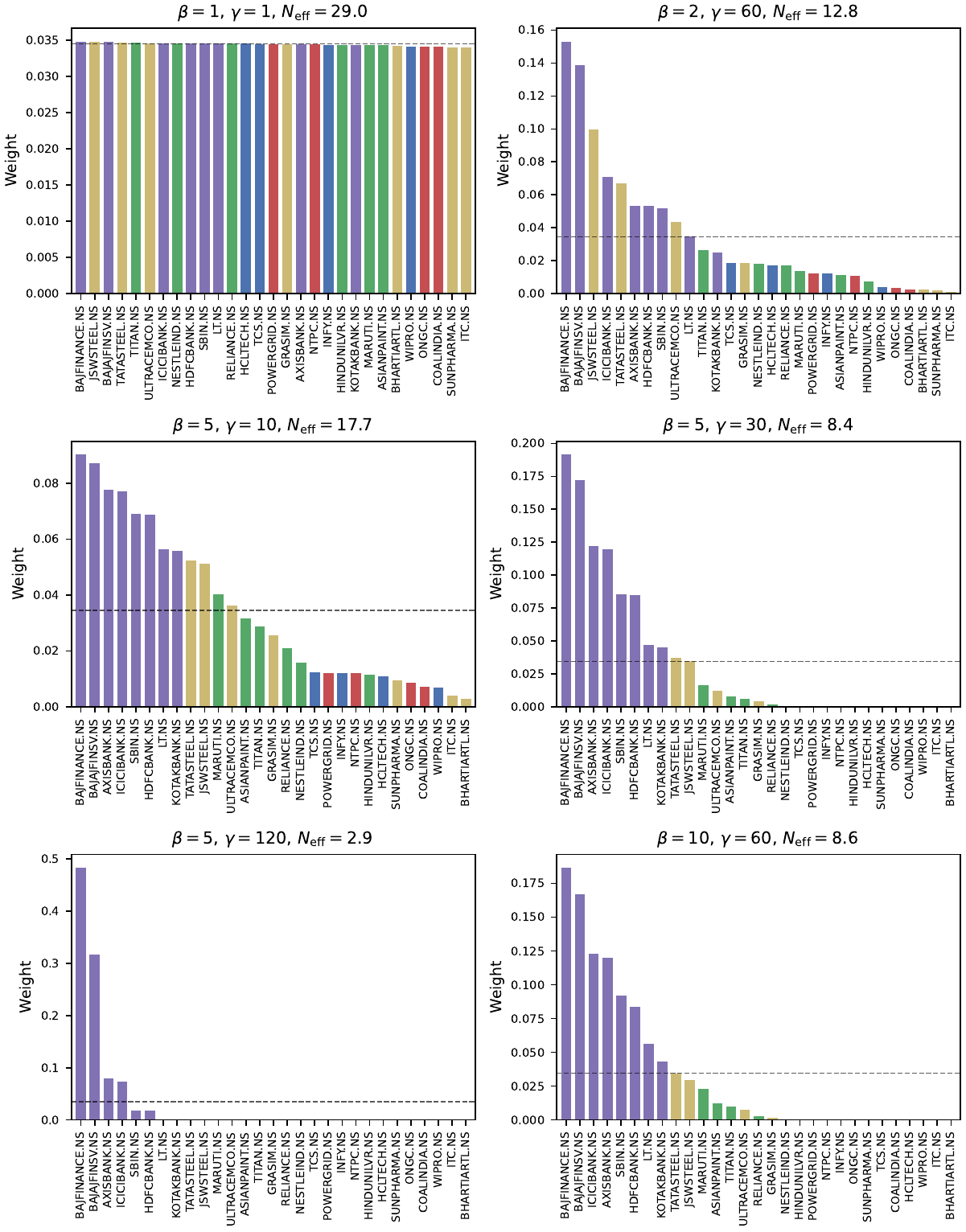}}
\caption[]{Figure~\ref{fig:weights} continued for India.}
\end{figure}

The \((1,1)\) allocations remain close to equal weighting. Larger \(\gamma\) generally
amplifies differences between scores, whereas \(\beta\) has a non-monotonic effect through the
magnetisations. At \((5,120)\), the three largest holdings account for 87.6\%, 96.1\%,
98.8\%, 95.1\%, and 88.3\% of capital in the US, UK, Germany, Japan, and India, respectively.
The histograms illustrate the range of attainable allocations without treating greater
concentration as preferable.

\FloatBarrier
\subsection{Risk--return characteristics of the XY portfolios}
Figure~\ref{fig:frontier} locates the six representative XY portfolios in mean--variance space.
Grey points denote random long-only portfolios, the black curve is the estimated efficient
frontier, and labelled coloured points identify the XY allocations.

\begin{figure}[p]
\centering
\makebox[\textwidth][c]{\includegraphics[width=1.08\textwidth]{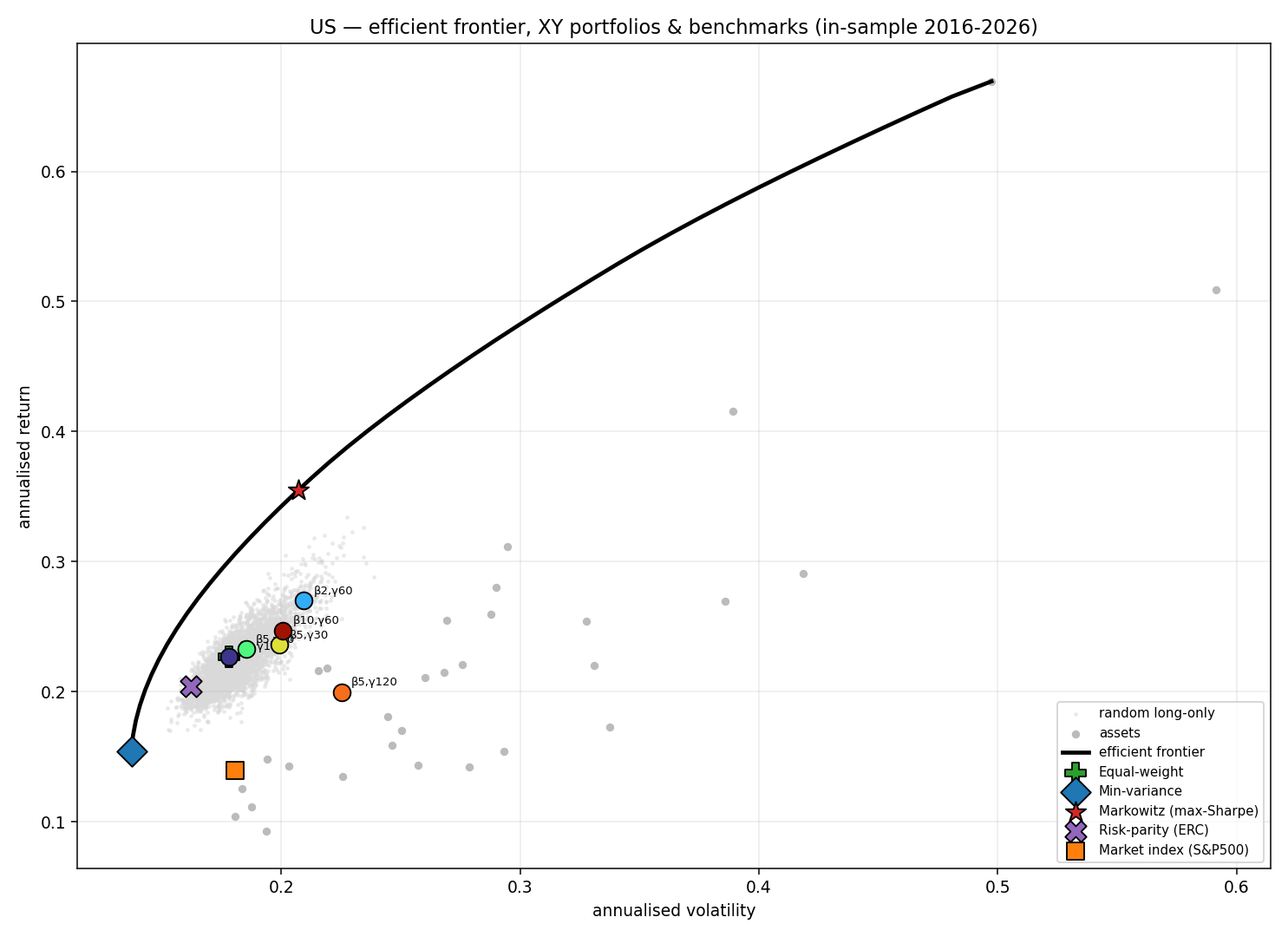}}
\caption{United States same-sample Markowitz frontier, XY portfolios, constituent assets, and
benchmarks.}
\label{fig:frontier}
\end{figure}
\clearpage

\begin{figure}[p]\ContinuedFloat
\centering
\makebox[\textwidth][c]{\includegraphics[width=1.08\textwidth]{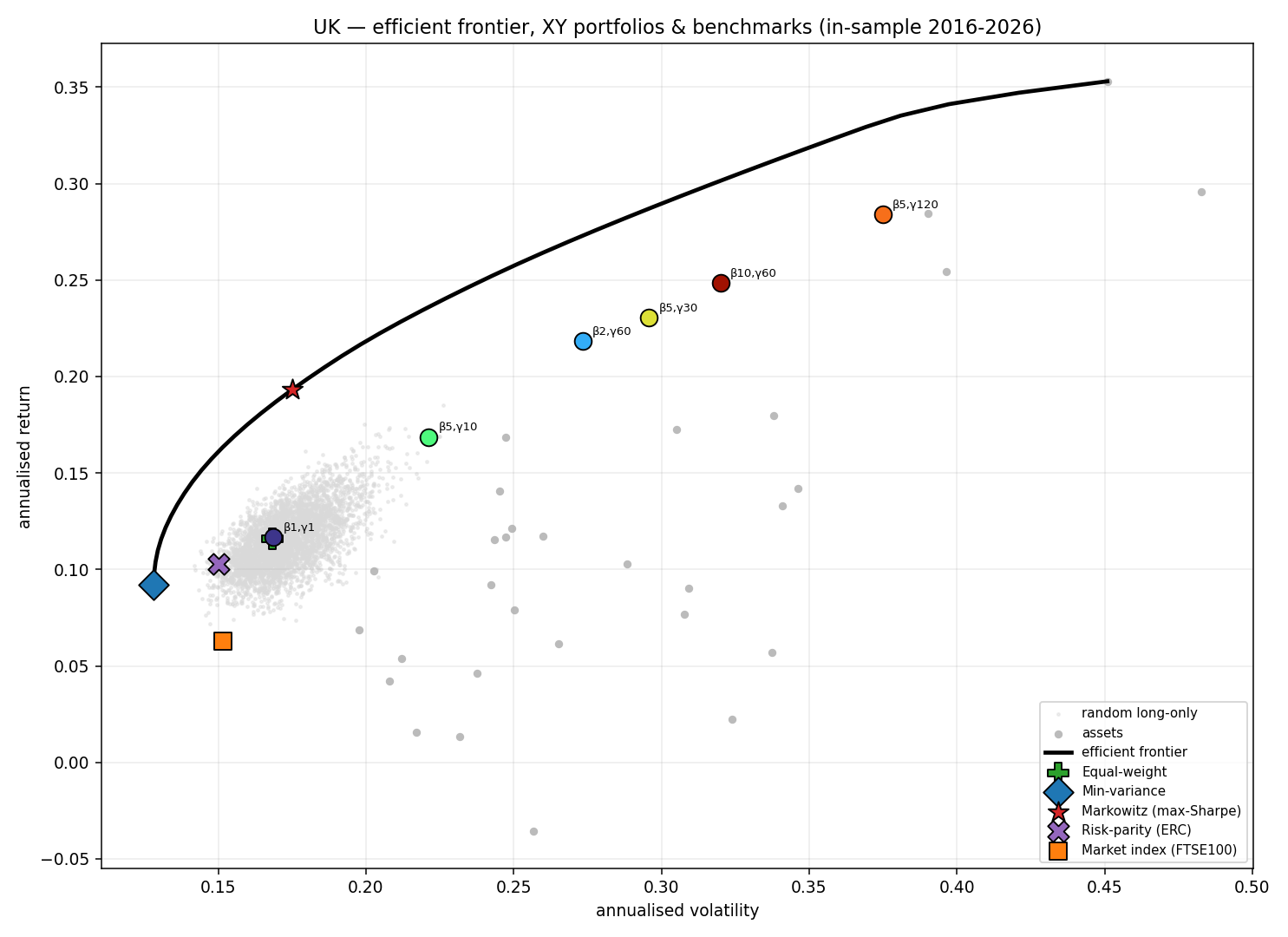}}
\caption[]{Figure~\ref{fig:frontier} continued for the United Kingdom.}
\end{figure}
\clearpage

\begin{figure}[p]\ContinuedFloat
\centering
\makebox[\textwidth][c]{\includegraphics[width=1.08\textwidth]{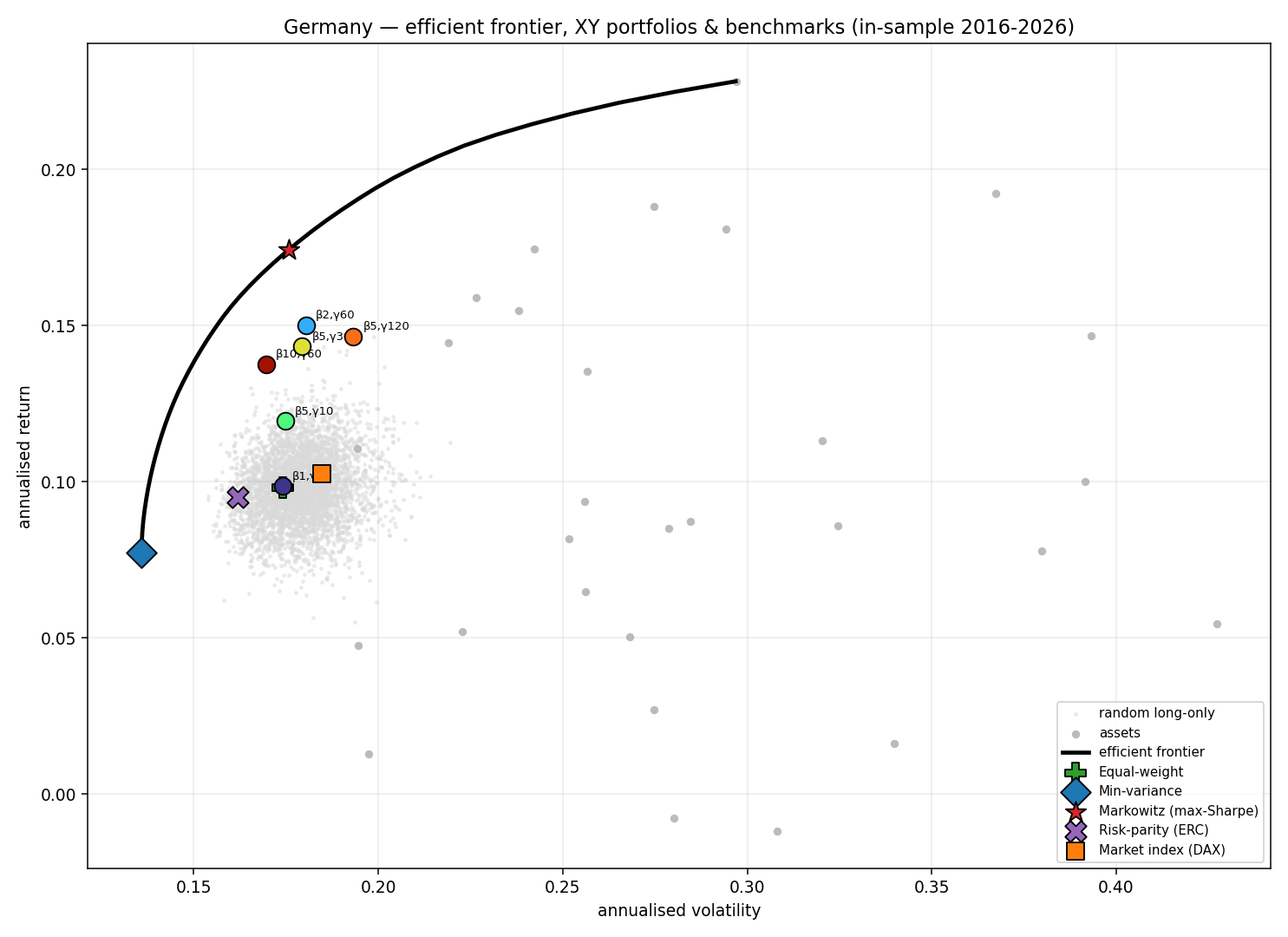}}
\caption[]{Figure~\ref{fig:frontier} continued for Germany.}
\end{figure}
\clearpage

\begin{figure}[p]\ContinuedFloat
\centering
\makebox[\textwidth][c]{\includegraphics[width=1.08\textwidth]{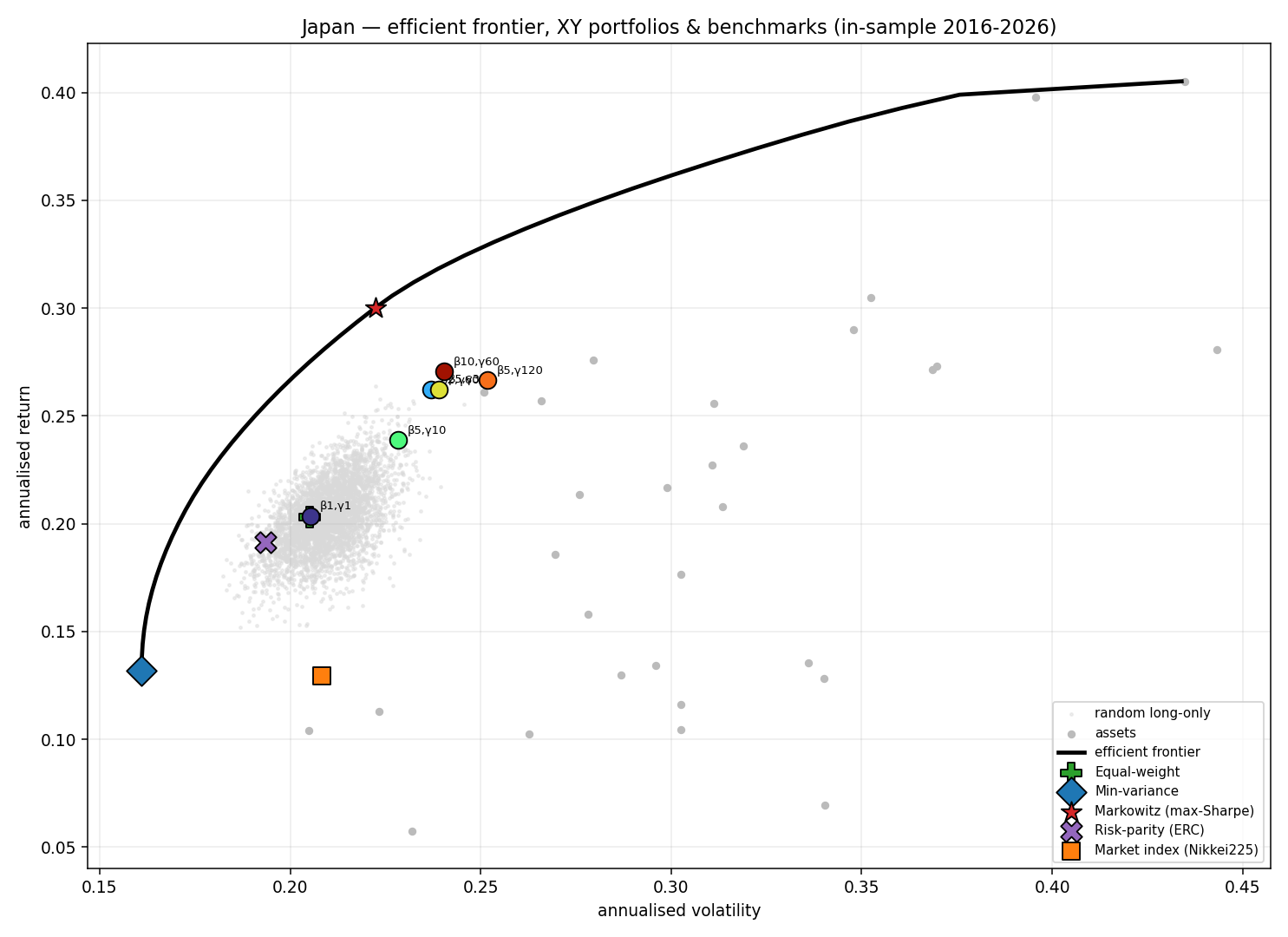}}
\caption[]{Figure~\ref{fig:frontier} continued for Japan.}
\end{figure}
\clearpage

\begin{figure}[p]\ContinuedFloat
\centering
\makebox[\textwidth][c]{\includegraphics[width=1.08\textwidth]{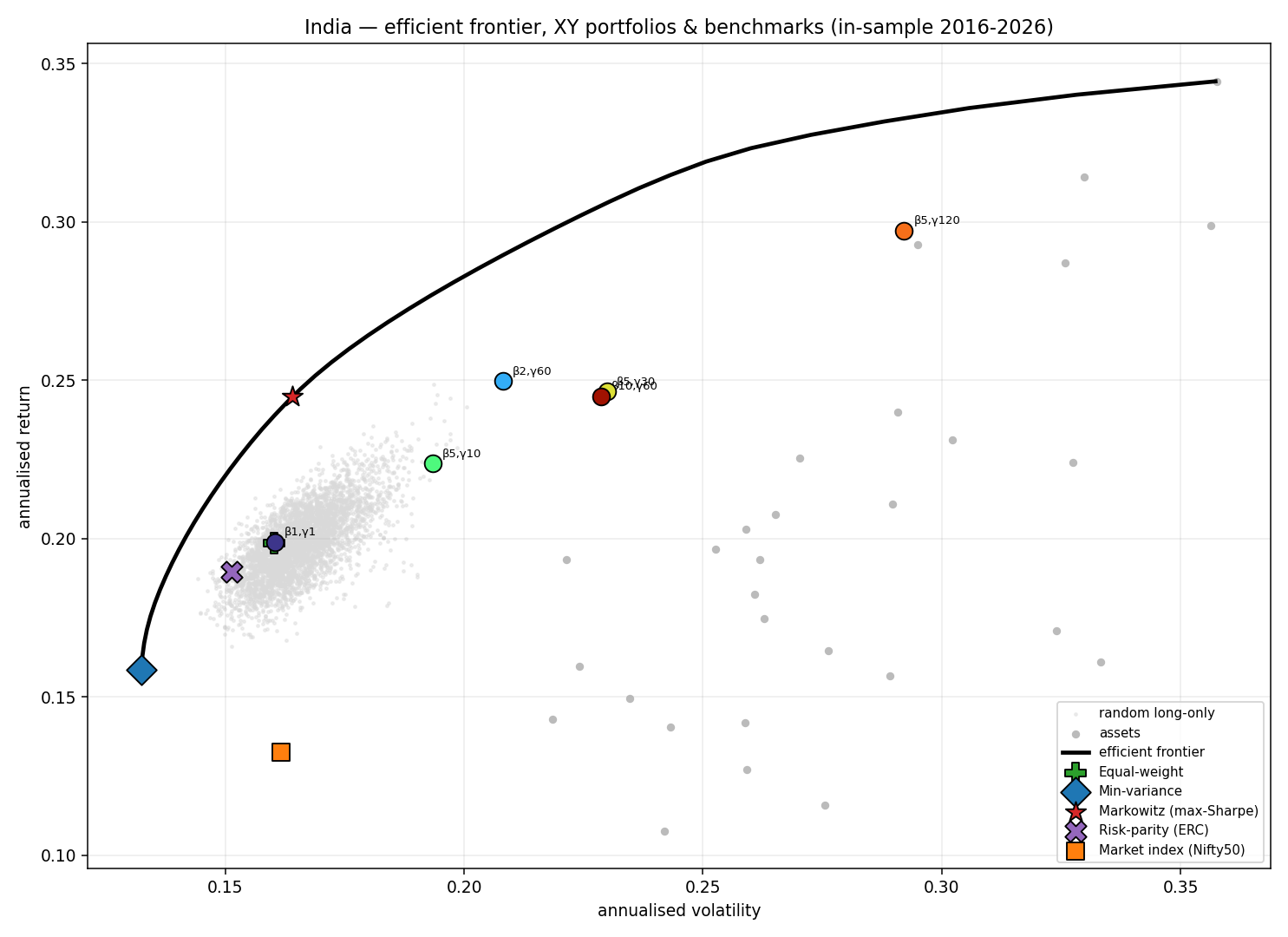}}
\caption[]{Figure~\ref{fig:frontier} continued for India.}
\end{figure}
\clearpage

Among the displayed XY points, the highest-Sharpe portfolios have annualised return and
volatility of 27.0\% and 20.9\% in the US, 21.8\% and 27.3\% in the UK, 15.0\% and 18.0\%
in Germany, 27.1\% and 24.0\% in Japan, and 19.9\% and 16.0\% in India. The score map does
not optimise a mean--variance objective. Its portfolios lie on or below the estimated efficient
boundary, and changing \((\beta,\gamma)\) moves them within the feasible set without tracing
that boundary.

\paragraph{Highest-Sharpe portfolios among the illustrative parameter pairs.}

Table~\ref{tab:bestxy} gives the ex-post maximum same-sample Sharpe ratio among the six manually
selected pairs. This comparison does not optimise over the continuous \((\beta,\gamma)\)
space or select parameters out of sample.

\begin{table}[htbp]
\centering
\small
\begin{tabular}{lccrrrr}
\toprule
Market & \(\beta\) & \(\gamma\) & Return (\%) & Volatility (\%) & Sharpe & \(\Neff\) \\
\midrule
US      & 2  & 60 & 27.0 & 20.9 & 1.289 & 15.07 \\
UK      & 2  & 60 & 21.8 & 27.3 & 0.798 &  7.83 \\
Germany & 2  & 60 & 15.0 & 18.0 & 0.832 &  8.43 \\
Japan   & 10 & 60 & 27.1 & 24.0 & 1.126 &  6.57 \\
India   & 1  &  1 & 19.9 & 16.0 & 1.240 & 29.00 \\
\bottomrule
\end{tabular}
\caption{Highest-Sharpe XY portfolio among the six illustrative parameter pairs in each market.
Returns and volatilities are annualised; \(\Neff\) is effective breadth.}
\label{tab:bestxy}
\end{table}

\FloatBarrier
\subsection{Comparison with standard benchmarks}
Table~\ref{tab:sharpe} compares these highest-Sharpe displayed XY portfolios with standard
long-only allocations evaluated on the same estimation sample.
\begin{table}[htbp]
\centering
\scriptsize
\setlength{\tabcolsep}{3.2pt}
\begin{tabular}{lcccccccc}
\toprule
Market & Highest XY & at \((\beta,\gamma)\) & \(\Neff\) & Equal-wt & Index & Min-var & ERC & Tangency \\
\midrule
US      & 1.289 & \((2,60)\)  & 15.07 & 1.273 & 0.772 & 1.124 & 1.257 & \textbf{1.713} \\
UK      & 0.798 & \((2,60)\)  &  7.83 & 0.689 & 0.415 & 0.718 & 0.685 & \textbf{1.104} \\
Germany & 0.832 & \((2,60)\)  &  8.43 & 0.563 & 0.556 & 0.569 & 0.587 & \textbf{0.992} \\
Japan   & 1.126 & \((10,60)\) &  6.57 & 0.989 & 0.622 & 0.819 & 0.989 & \textbf{1.350} \\
India   & 1.240 & \((1,1)\)   & 29.00 & 1.239 & 0.820 & 1.197 & 1.252 & \textbf{1.492} \\
\bottomrule
\end{tabular}
\caption{Same-sample Sharpe ratios. ``Highest XY'' is the ex-post maximum among the six displayed
parameter pairs. ERC denotes equal-risk contribution.}
\label{tab:sharpe}
\end{table}

In the United States, United Kingdom, Germany, and Japan, the highest-Sharpe displayed XY
portfolio exceeds equal weighting, the headline index, minimum variance, and equal-risk
contribution on this sample. India's value of 1.240 is essentially equal to the equal-weight
ratio of 1.239 and below the equal-risk-contribution ratio of 1.252. The tangency portfolio has
the largest ratio in every market, as expected from its explicit maximisation of the same
in-sample criterion.

Figure~\ref{fig:bench} separates annualised arithmetic return from Sharpe ratio. Grey bands show
the 5th--95th percentiles of 4000 random Dirichlet portfolios, providing reference ranges rather
than confidence intervals.

\begin{figure}[p]
\centering
\makebox[\textwidth][c]{\includegraphics[width=1.08\textwidth]{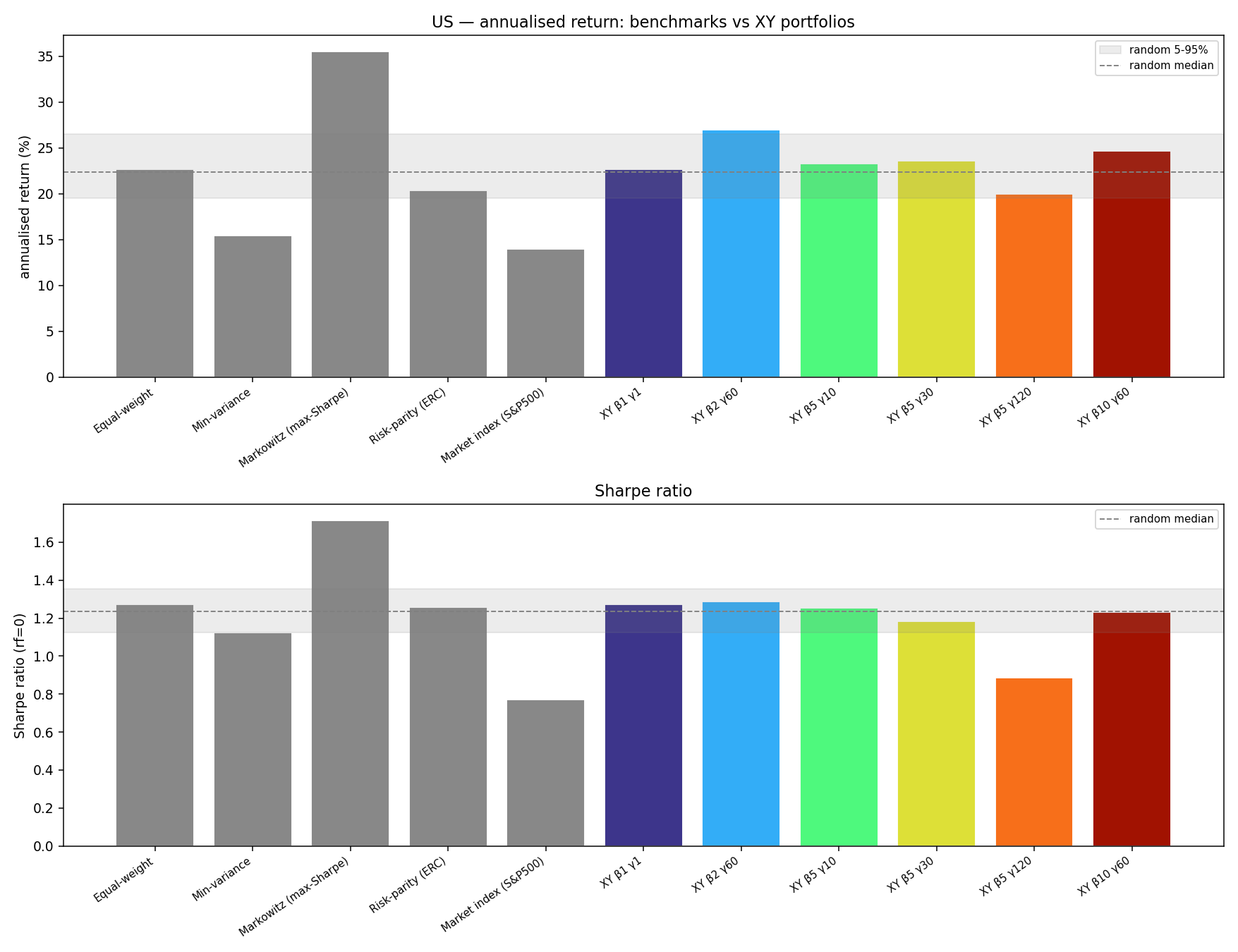}}
\caption{United States annualised returns (upper panel) and zero-rate Sharpe ratios (lower panel)
for benchmarks and the six XY portfolios.}
\label{fig:bench}
\end{figure}

\begin{figure}[p]\ContinuedFloat
\centering
\makebox[\textwidth][c]{\includegraphics[width=1.08\textwidth]{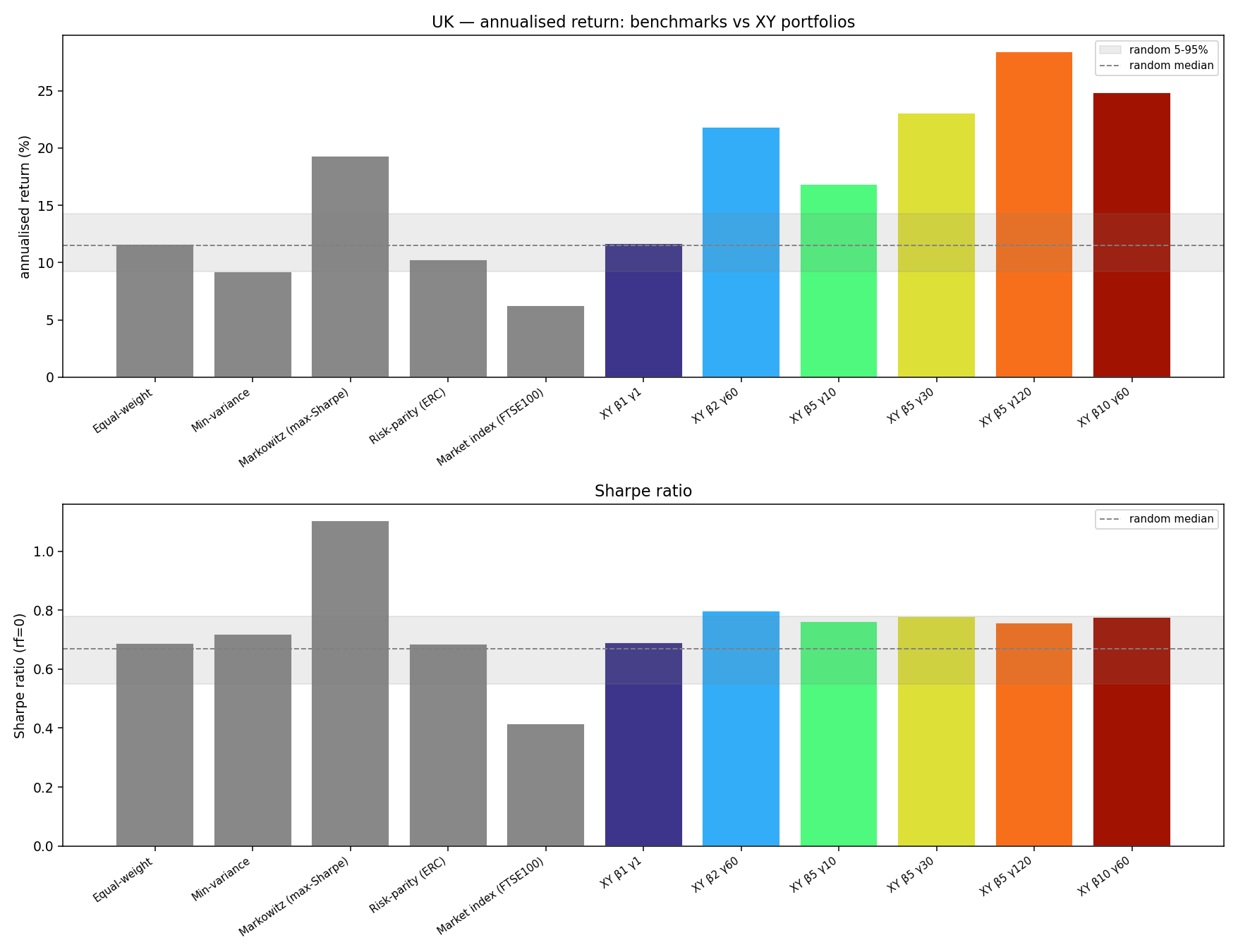}}
\caption[]{Figure~\ref{fig:bench} continued for the United Kingdom.}
\end{figure}

\begin{figure}[p]\ContinuedFloat
\centering
\makebox[\textwidth][c]{\includegraphics[width=1.08\textwidth]{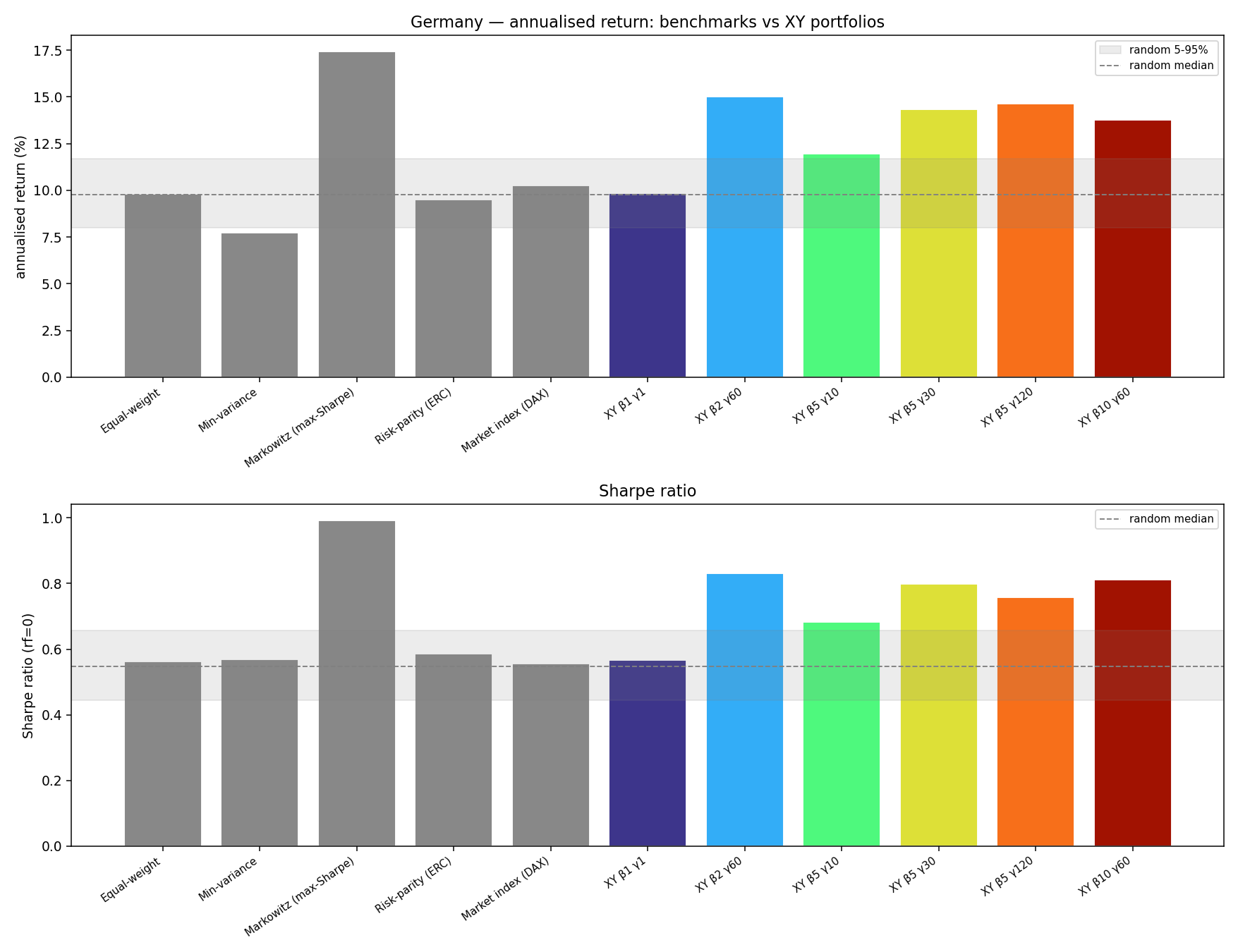}}
\caption[]{Figure~\ref{fig:bench} continued for Germany.}
\end{figure}

\begin{figure}[p]\ContinuedFloat
\centering
\makebox[\textwidth][c]{\includegraphics[width=1.08\textwidth]{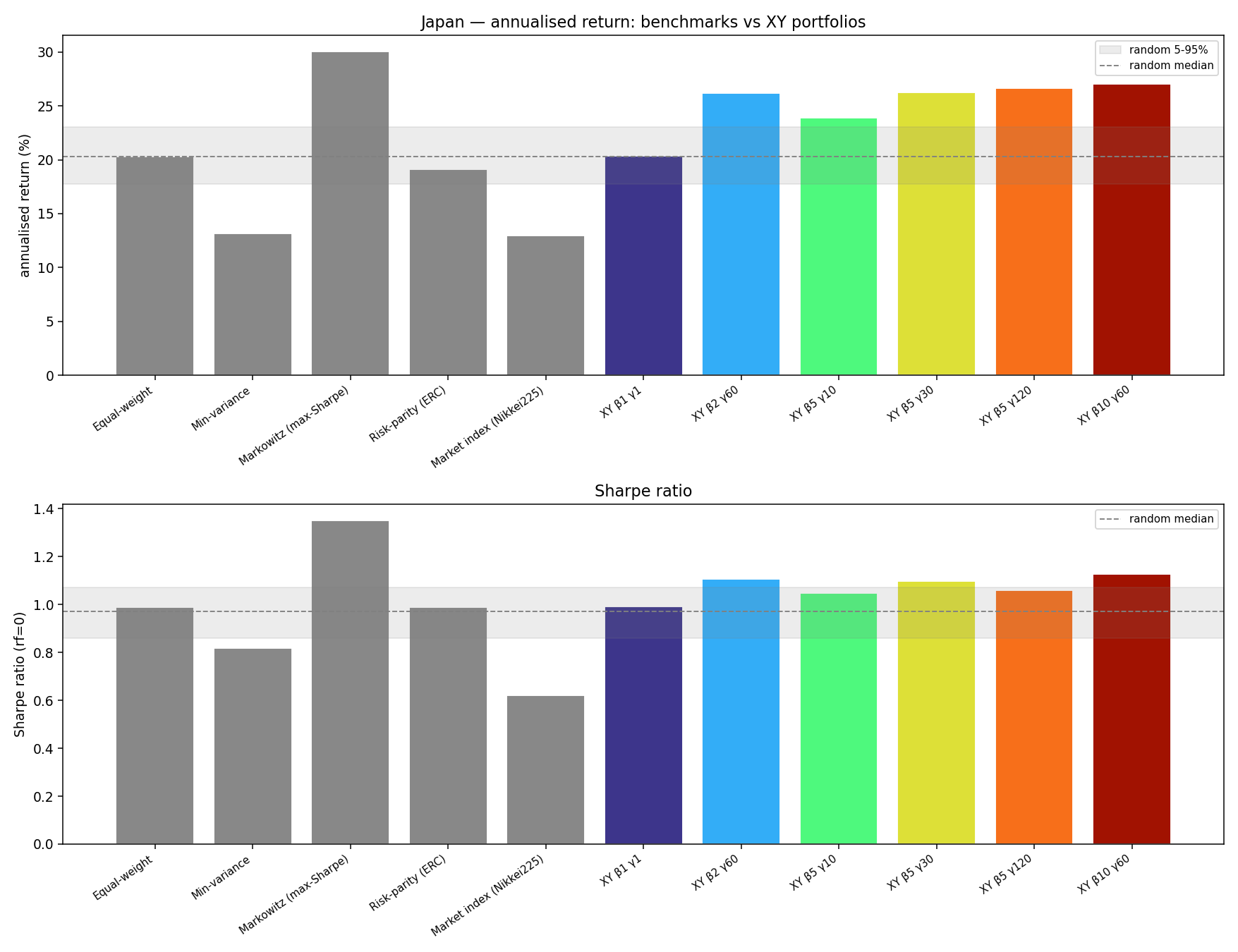}}
\caption[]{Figure~\ref{fig:bench} continued for Japan.}
\end{figure}

\begin{figure}[p]\ContinuedFloat
\centering
\makebox[\textwidth][c]{\includegraphics[width=1.08\textwidth]{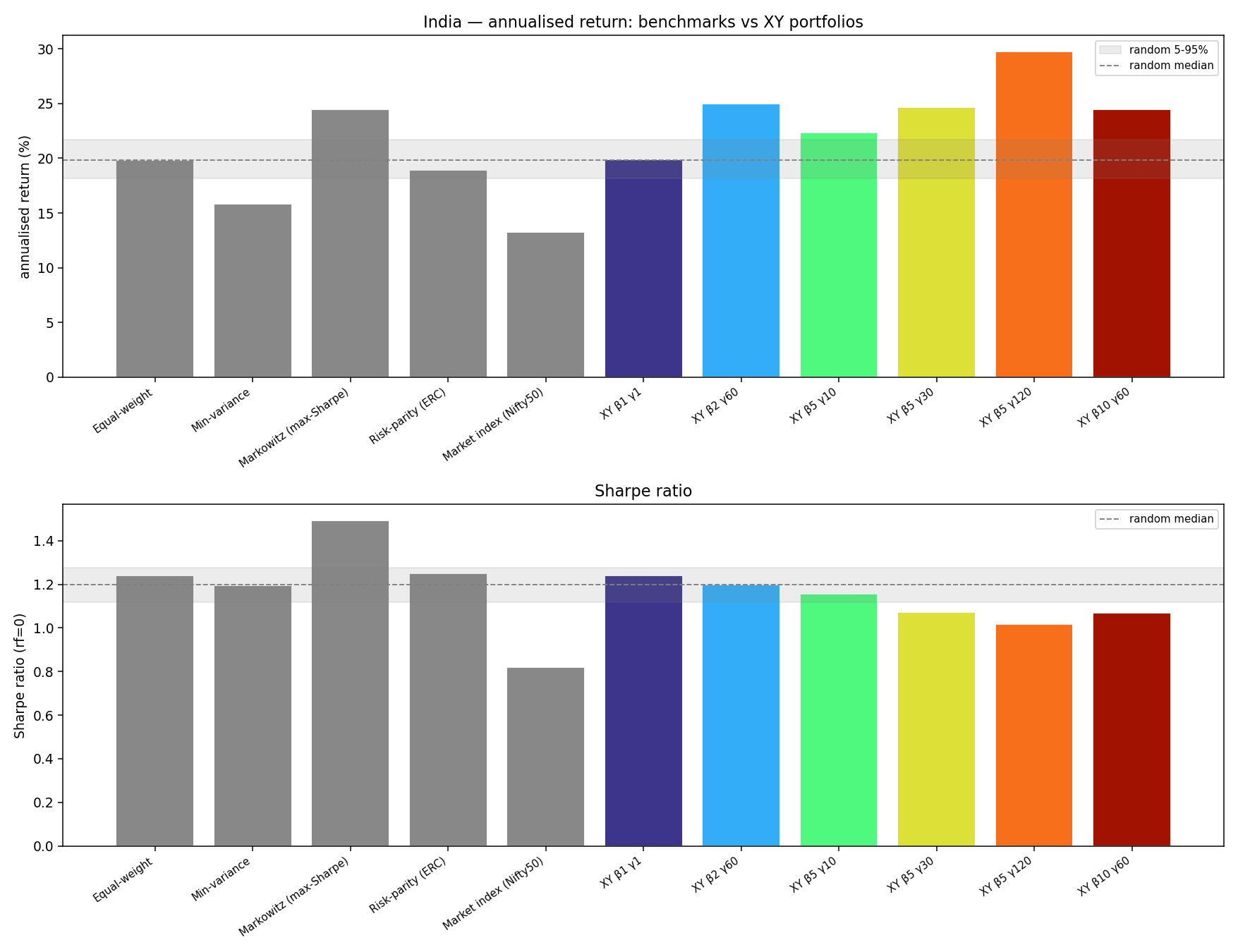}}
\caption[]{Figure~\ref{fig:bench} continued for India.}
\end{figure}
\clearpage

Return and Sharpe ratio can favour different parameter pairs. In the UK, \((5,120)\) gives the
largest displayed XY return, 28.4\%, but volatility of 37.5\% reduces its Sharpe ratio to 0.758;
\((2,60)\) has the higher ratio of 0.798. In India, \((5,120)\) returns 29.7\% with Sharpe
1.018, below the near-equal \((1,1)\) portfolio's 1.240.

\subsection{Synthesis of the numerical results}
The geometric diagnostics and Ward ordering produce paths that keep relatively strong local
pairs while retaining only a small fraction of total absolute off-diagonal correlation. Cutoff,
normalisation, and response checks establish the numerical consistency of contraction on those
paths. Across \((\beta,\gamma)\), the allocations range from nearly equal weighting to strong
concentration, with non-monotonic dependence on \(\beta\) at large \(\gamma\). Several
representative portfolios have same-sample Sharpe ratios comparable with or above non-tangency
benchmarks. These findings establish computational feasibility and describe model behaviour;
predictive investment performance remains untested.

\section{Discussion}
\label{sec:discussion}
\subsection{Interpretation of the tensor-network construction}
The tensor-network method evaluates a continuous-spin model through a finite integer-current
representation. On the selected path, the bottom-up and top-down sweeps require \(O(ND^2)\)
operations for \(D=2K+1\). Their marginals give \(m_i=\langle\cos\theta_i\rangle\), and
the softmax converts these scores into feasible long-only weights. This separation also explains
the roles of the parameters: \(\beta\) controls the Gibbs response to fields and couplings,
while \(\gamma\) controls how score differences translate into concentration. Together they
generate a continuous family of allocations.

\subsection{Empirical interpretation and scope}
The five-market study indicates approximate tree geometry, but the Ward paths retain only a
small fraction of the dense correlations. The parameter surfaces and weight profiles span
near-equal and highly concentrated allocations; the six displayed \((\beta,\gamma)\) pairs
illustrate this continuum. Their positions in the long-only feasible set need not follow the
Markowitz frontier because the XY scoring construction uses a different objective.

\section{Conclusion}
\label{sec:conclusion}

We have applied tensor-network contraction to an interacting XY model for long-only portfolio
construction. Local fields encode expected-return information, pairwise couplings encode
empirical dependence, and equilibrium spin projections supply asset scores. A final softmax gives
positive capital weights that sum to one. The angles remain latent variables throughout the
inference; allocation occurs only when the scores are mapped to weights.

Several displayed portfolios have favourable same-sample Sharpe ratios relative to equal
weighting, the headline index, minimum variance, and equal-risk contribution. The Markowitz
tangency portfolio remains strongest because it maximises the evaluation criterion on that same
sample. The results demonstrate computational feasibility and parameter-dependent allocation
behaviour. Their interpretation is limited by current truncation, replacement of the dense graph
with a Ward-ordered path, same-sample parameter inspection, omitted trading costs, and the absence
of rolling out-of-sample tests.

\section*{Acknowledgements}
Kartikeya Chowdhry acknowledges the use of Claude Opus 4.8 for assistance in writing code used in this study.
\section*{Declaration of competing interest}
The authors declare that they have no known competing financial interests or personal relationships that could have appeared to influence the work reported in this paper.
\appendix
\section{Tensor-network architectures and contraction complexity}
\label{app:tensor-background}

\begin{figure}[htbp]
\centering
\begin{subfigure}[b]{0.31\textwidth}
\centering
\begin{tikzpicture}[
  x=0.58cm,y=0.58cm,
  tn/.style={draw,circle,fill=MidnightBlue!12,inner sep=1.8pt},
  every path/.style={line width=0.45pt}]
  \foreach \x in {0,...,4}{
    \node[tn] (m\x) at (\x,0.8) {};
    \draw (m\x) -- ++(0,-0.55);
  }
  \foreach \x in {0,...,3}{
    \pgfmathtruncatemacro{\y}{\x+1}
    \draw (m\x) -- (m\y);
  }
\end{tikzpicture}
\caption{MPS}
\end{subfigure}\hfill
\begin{subfigure}[b]{0.31\textwidth}
\centering
\begin{tikzpicture}[
  x=0.55cm,y=0.55cm,
  tn/.style={draw,circle,fill=ForestGreen!14,inner sep=1.55pt},
  every path/.style={line width=0.42pt}]
  \foreach \x in {0,1,2}{
    \foreach \y in {0,1,2}{
      \node[tn] (p\x\y) at (\x,\y) {};
      \draw (p\x\y) -- ++(-0.28,-0.28);
    }
  }
  \foreach \y in {0,1,2}{
    \draw (p0\y)--(p1\y)--(p2\y);
  }
  \foreach \x in {0,1,2}{
    \draw (p\x0)--(p\x1)--(p\x2);
  }
\end{tikzpicture}
\caption{PEPS}
\end{subfigure}\hfill
\begin{subfigure}[b]{0.34\textwidth}
\centering
\begin{tikzpicture}[
  x=0.42cm,y=0.55cm,
  iso/.style={draw,circle,fill=MidnightBlue!12,inner sep=1.35pt},
  dis/.style={draw,rectangle,fill=BrickRed!14,inner sep=1.2pt,font=\tiny},
  every path/.style={line width=0.42pt}]
  \foreach \x in {0,...,7}{
    \coordinate (q\x) at (\x,0);
    \draw (q\x)--++(0,-0.32);
  }
  \node[iso] (w0) at (0.5,0.8) {};
  \node[iso] (w1) at (2.5,0.8) {};
  \node[iso] (w2) at (4.5,0.8) {};
  \node[iso] (w3) at (6.5,0.8) {};
  \draw (q0)--(w0)--(q1) (q2)--(w1)--(q3)
        (q4)--(w2)--(q5) (q6)--(w3)--(q7);
  \node[dis] (u0) at (1.5,0.27) {\(u\)};
  \node[dis] (u1) at (5.5,0.27) {\(u\)};
  \draw (q1)--(u0)--(q2) (q5)--(u1)--(q6);
  \node[iso] (v0) at (1.5,1.65) {};
  \node[iso] (v1) at (5.5,1.65) {};
  \draw (w0)--(v0)--(w1) (w2)--(v1)--(w3);
  \node[dis] (u2) at (3.5,1.2) {\(u\)};
  \draw (w1)--(u2)--(w2);
  \node[iso] (r) at (3.5,2.45) {};
  \draw (v0)--(r)--(v1);
\end{tikzpicture}
\caption{MERA}
\end{subfigure}
\caption{Schematic tensor-network geometries. Nodes are tensors, internal lines are contracted
bonds, and dangling lines are open indices. In MERA, \(u\) denotes a disentangler.}
\label{fig:tnarchitectures}
\end{figure}

Figure~\ref{fig:tnarchitectures} illustrates three standard tensor-network architectures
\cite{orus2014,biamonte2017}. A matrix product state (MPS) factorises a one-dimensional object
along a chain and admits sequential contraction sweeps. A projected entangled-pair state (PEPS)
extends this construction to higher-dimensional lattices, retaining spatial locality at the cost
of loops. A multiscale entanglement-renormalisation ansatz (MERA) alternates disentanglers with
coarse-graining tensors, giving a hierarchy with narrow causal cones and access to several length
scales.

These architectures are useful in quantum many-body physics partly because they accommodate
\emph{area laws}. For many low-energy local quantum states, the entanglement entropy of a region
\(A\) grows with its boundary \(|\partial A|\) rather than the volume of \(A\). A cut through a tensor
network gives the bound
\begin{equation}
    S(A)\ \lesssim\ \sum_{e\in\operatorname{cut}(A)}\log\chi_e .
\label{eq:arealaw}
\end{equation}
An MPS cut through a half-chain crosses one bond and supports bounded entropy at fixed \(\chi\).
A PEPS cut crosses a number of bonds proportional to the boundary, naturally accommodating an
area law in higher dimensions. Scale-dependent bonds allow MERA to represent the logarithmic
corrections of one-dimensional critical systems. In this paper the network represents a
classical partition function; the entanglement interpretation provides background only.

Topology also governs contraction complexity \cite{ran2020}. Standard MPS observables have
polynomial cost in system size, typically \(O(Nd\chi^3)\). Exact PEPS contraction is generally
exponential in lattice width and requires approximations. Local MERA observables involve a
bounded-width causal cone across \(O(\log N)\) layers, each with a model-dependent polynomial
cost in \(\chi\). The Ward-ordered path used here is MPS-like: truncating the Fourier--Bessel
current to \(n\in\{-K,\ldots,K\}\) gives \(D=2K+1\), and the specialised scalar messages
cost \(O(ND^2)\). At fixed path and cutoff, contraction is exact up to floating-point
arithmetic. Current truncation and replacement of the dense graph by a sparse graph remain
separate approximations.

\section{Implementation scales and asset universes}
\label{app:implementation}

\subsection{Scale and cutoff diagnostics}
Table~\ref{tab:implementation} records the largest retained path coupling, the range of daily
fields, and the adaptive current window at the largest inverse temperature among the six
illustrative parameter pairs. These values specify the scales of the empirical Gibbs models.

\begin{table}[htbp]
\centering
\small
\begin{tabular}{lrrrr}
\toprule
Market & \(J_{\max}\) & \(\min_i h_i\) & \(\max_i h_i\) & \(K(10)\) \\
\midrule
US      & 0.8991 &  0.02139 & 0.06959 & 40 \\
UK      & 0.8764 & -0.01682 & 0.03521 & 39 \\
Germany & 0.7843 & -0.01213 & 0.03876 & 37 \\
Japan   & 0.9184 &  0.00199 & 0.05760 & 40 \\
India   & 0.7599 &  0.01732 & 0.05361 & 37 \\
\bottomrule
\end{tabular}
\caption{Empirical parameter scales on the Ward-ordered paths. Fields use the daily
\(\hat\mu_i/\hat\sigma_i\) convention with \(\eta=1\), and \(K(10)\) follows
Equation~\eqref{eq:adaptiveK}.}
\label{tab:implementation}
\end{table}

\subsection{Retained asset lists}
The following lists are the exact columns of the complete-case matrices used for estimation.

\begin{sloppypar}\small\raggedright
\noindent\textbf{United States (30).}
\texttt{AAPL}, \texttt{MSFT}, \texttt{GOOGL}, \texttt{AMZN}, \texttt{NVDA}, \texttt{META},
\texttt{TSLA}, \texttt{BRK-B}, \texttt{JPM}, \texttt{V}, \texttt{JNJ}, \texttt{WMT},
\texttt{MA}, \texttt{PG}, \texttt{HD}, \texttt{AVGO}, \texttt{ORCL}, \texttt{CVX},
\texttt{XOM}, \texttt{LLY}, \texttt{ABBV}, \texttt{KO}, \texttt{PEP}, \texttt{COST},
\texttt{MRK}, \texttt{ADBE}, \texttt{CSCO}, \texttt{ACN}, \texttt{MCD}, \texttt{NFLX}.

\medskip
\noindent\textbf{United Kingdom (30).}
\texttt{AZN.L}, \texttt{SHEL.L}, \texttt{HSBA.L}, \texttt{ULVR.L}, \texttt{BP.L},
\texttt{GSK.L}, \texttt{DGE.L}, \texttt{RIO.L}, \texttt{GLEN.L}, \texttt{BATS.L},
\texttt{REL.L}, \texttt{NG.L}, \texttt{VOD.L}, \texttt{BARC.L}, \texttt{LLOY.L},
\texttt{NWG.L}, \texttt{PRU.L}, \texttt{AAL.L}, \texttt{TSCO.L}, \texttt{BA.L},
\texttt{RR.L}, \texttt{IMB.L}, \texttt{CPG.L}, \texttt{EXPN.L}, \texttt{AV.L},
\texttt{STAN.L}, \texttt{ANTO.L}, \texttt{SSE.L}, \texttt{WTB.L}, \texttt{SGRO.L}.

\medskip
\noindent\textbf{Germany (29).}
\texttt{SAP.DE}, \texttt{SIE.DE}, \texttt{ALV.DE}, \texttt{DTE.DE}, \texttt{BMW.DE},
\texttt{BAS.DE}, \texttt{BAYN.DE}, \texttt{IFX.DE}, \texttt{VOW3.DE}, \texttt{ADS.DE},
\texttt{DB1.DE}, \texttt{MUV2.DE}, \texttt{RWE.DE}, \texttt{EOAN.DE}, \texttt{MRK.DE},
\texttt{HEN3.DE}, \texttt{BEI.DE}, \texttt{FRE.DE}, \texttt{CON.DE}, \texttt{DBK.DE},
\texttt{HEI.DE}, \texttt{SY1.DE}, \texttt{VNA.DE}, \texttt{ZAL.DE}, \texttt{PUM.DE},
\texttt{FME.DE}, \texttt{BNR.DE}, \texttt{QIA.DE}, \texttt{LHA.DE}.

\medskip
\noindent\textbf{Japan (30).}
\texttt{7203.T}, \texttt{6758.T}, \texttt{6861.T}, \texttt{8306.T}, \texttt{9984.T},
\texttt{6098.T}, \texttt{9432.T}, \texttt{8035.T}, \texttt{6501.T}, \texttt{7974.T},
\texttt{4063.T}, \texttt{6902.T}, \texttt{8058.T}, \texttt{9433.T}, \texttt{6594.T},
\texttt{4502.T}, \texttt{7267.T}, \texttt{8316.T}, \texttt{8001.T}, \texttt{6981.T},
\texttt{6367.T}, \texttt{4519.T}, \texttt{8411.T}, \texttt{7741.T}, \texttt{6273.T},
\texttt{6146.T}, \texttt{4661.T}, \texttt{8031.T}, \texttt{7011.T}, \texttt{4543.T}.

\medskip
\noindent\textbf{India (29).}
\texttt{RELIANCE.NS}, \texttt{TCS.NS}, \texttt{HDFCBANK.NS}, \texttt{INFY.NS},
\texttt{ICICIBANK.NS}, \texttt{HINDUNILVR.NS}, \texttt{ITC.NS}, \texttt{SBIN.NS},
\texttt{BHARTIARTL.NS}, \texttt{KOTAKBANK.NS}, \texttt{LT.NS}, \texttt{HCLTECH.NS},
\texttt{AXISBANK.NS}, \texttt{BAJFINANCE.NS}, \texttt{ASIANPAINT.NS}, \texttt{MARUTI.NS},
\texttt{SUNPHARMA.NS}, \texttt{TITAN.NS}, \texttt{WIPRO.NS}, \texttt{ULTRACEMCO.NS},
\texttt{NESTLEIND.NS}, \texttt{ONGC.NS}, \texttt{NTPC.NS}, \texttt{POWERGRID.NS},
\texttt{TATASTEEL.NS}, \texttt{JSWSTEEL.NS}, \texttt{COALINDIA.NS},
\texttt{BAJAJFINSV.NS}, \texttt{GRASIM.NS}.
\end{sloppypar}

\end{document}